\documentclass[12pt,preprint]{aastex}

\slugcomment{}

\shorttitle{Extremely Light Bosonic Dark Matter}
\shortauthors{Woo \& Chiueh}

\begin{document}

\title{High-Resolution Simulation on Structure Formation with Extremely Light Bosonic Dark Matter}

\author{Tak-Pong Woo\altaffilmark{1,3} and Tzihong Chiueh\altaffilmark{1,2,3}}

\email{bonwood@scu.edu.tw \& chiuehth@phys.ntu.edu.tw}

\altaffiltext{1}{Department of Physics, National Taiwan University, 106, Taipei, Taiwan, R.O.C.}
\altaffiltext{2}{Center for Theoretical Sciences, National Taiwan University, 106, Taipei, Taiwan, R.O.C. }
\altaffiltext{3}{LeCosPa, National Taiwan University, 106, Taipei, Taiwan, R.O.C.}

\begin{abstract}
A bosonic dark matter model is examined in detail via high-resolution simulations.
These bosons have particle mass of order $10^{-22}eV$ and are non-interacting. If they do exist and can
account for structure formation, these bosons must be condensed into the Bose-Einstein state and
described by a coherent wave function.  This matter, also known as \emph{Fuzzy Dark Matter} \citep{hu00}, is speculated
to be able, first, to eliminate the sub-galactic halos to solve the problem of over-abundance of dwarf
galaxies, and, second, to produce flat halo cores in galaxies suggested by some observations.  We investigate this model with simulations up to $1024^3$ resolution
in an 1 $h^{-1}Mpc$ box that maintains the background matter density $\Omega_m=0.3$ and $\Omega_\Lambda=0.7$.
Our results show that the extremely light bosonic dark matter (ELBDM) can indeed eliminate low-mass halos through the suppression of short-wavelength fluctuations, as predicted by the linear perturbation theory.
But to the contrary of expectation, our simulations yield singular cores in the collapsed halos, where the halo density profile is similar, but not identical, to the NFW profile \citep{nfw97}. Such a profile arises regardless of whether the halo forms through accretion or merger.  In addition, the virialized halos exhibit anisotropic turbulence inside a well-defined virial boundary.  Much like the velocity dispersion of standard dark matter particles, turbulence is dominated by the random radial flow in most part of the halos and becomes isotropic toward the halo cores.  Consequently the three-dimensional collapsed halo mass distribution can deviate from spherical symmetry, as the cold dark matter halo does.
\end{abstract}

\keywords{cosmology, dark matter, galaxies, large-scale structure
of universe}

\section{Introduction}

Observations of low surface brightness galaxies and dwarf galaxies
indicate that the cores of galactic halos have shallow density
profiles \citep{dal97,swt00} instead of central cusps
predicted by cold dark matter (CDM) \citep{nfw97}.
In addition, the number density of dwarf
galaxies in Local Group turns out to be an order of magnitude fewer
than that produced by CDM simulations \citep{kkvp99}.  These two features cast doubt on the validity of
standard CDM.  There have been at least three different solutions
proposed to resolve these problems: (1) warm dark matter, (2)
collisional dark matter and (3) fuzzy dark matter.

Warm dark matter can suppress small-scale structures by free
streaming.  It seems to be able to both solve
the over-abundance problem of dwarf galaxies and the singular core problem.
In this model the flat core is embedded within a radius a couple of percents of the virial radius \citep{jing01,colins08}, and the core smoothly connects to the NFW profile \citep{nfw97} outside.
However, this modification may generally adversely affect structures of
somewhat larger scales \citep{hu00}, despite that fine tuning of
the thermal velocity of dark matter particles may still be able to have
the larger scale structures consistent with observations \citep{abazajian06, viel08}.

For collisional dark matter, the halo core can be flattened and dwarf
galaxies destroyed, and N-body simulations
confirm this conjecture \citep{spst00}.  But simulations
also show that very frequent collisions can yield even more
singular cores than the standard collisionless CDM does \citep{yoshida00}. This opposite behavior is indicative of
the requirement of fine tuning for collisional parameters.

The third solution to the problem is to treat dark matter as an
\emph{extremely light bosonic dark matter} (ELBDM) or \emph{fuzzy dark matter} \citep{pr90,sin94,hu00}.
Axion has been thought to be a candidate of light bosonic dark
matter. But for the light dark matter to erase the singular galactic core
and suppresses low-mass halos, the particle mass must be far smaller than axion (m$\sim$ $10^{-22}$ eV), so low that the uncertainty
principle operates on the astronomical length scale.  Much like
axions, the ELBDM is in a Bose-Einstein condense state produced in
the early universe. These extremely light particles share the common
ground state and is described by a single coherent wave function. Its
de-Broglie wavelength is comparable to or even somewhat smaller than the
Jean's length \citep{dw97}, where the quantum
fluctuation provides effective pressure against self-gravity.

Several previous works have pondered on such an idea or its variance \citep{sin94,hu00,slopez03}, in which the wave mechanics
is described by the Schr\"{o}dinger-Poisson equation with Newtonian
gravity or by the Klein-Gordon equation with gravity.  The Schr\"{o}dinger-Poisson system addresses the scale-free regime of quantum mechanics, where the Jean's length is a dynamical running parameter.  On the other hand, the Klein-Gordon system makes use of the Compton wavelength as a natural length scale to create the flat core in a halo.  Widrow \& Kaiser conducted simulations for the two-dimensional Schr\"{o}dinger-Poisson system to approximate the standard
collisionless cold dark matter \citep{wk93}.  In the 2D case, the $1/r$ gravitational potential is replaced by $log(r)$, and
the 2D force law in their simulation becomes of longer range than it
actually is in 3D.  Due to the lack of 3D numerical simulations,
some authors resort to spherical symmetry \citep{sin94,slopez03} or even 1D \citep{hu00} to study this
problem. These simplifications may not capture what actually results
in a 3D system with realistic initial conditions. In particular, the
existence of a flattened core has been derived or inferred from these
previous works of 1D system or with spherical symmetry.  In this paper we report high-resolution fully 3D
simulations for this problem. Surprisingly, our simulations
reveal that the singular cores of bound objects remain to exist even when
the core size is much smaller than the Jean's length.

In Sec.2, we provide an explanation for the possible existence
of the Bose-Einstein state for the extremely low mass bosons under
investigation here. We then discuss two different representations of
ELBDM and the evolution of linear perturbations for the two representations.
In Sec.3, the numerical
scheme and initial condition are described. We present the simulation results in Sec.4.
In Sec.5, we look into the physics of collapsed cores with detailed analyses
from different perspectives.  Finally the conclusion is given in Sec.6.
In the Appendix, we present results of 1D and 2D simulations and demonstrate singular cores do not occur in 1D and 2D cases.

\section{Theory}
\subsection{Bose-Einstein Condensate}

A Bose-Einstein condensate (BEC) is a state of bosons cooled to a
temperature below the critical temperature. BEC happens after a
phase transition where a large fraction of particles condense into
the ground state, at which point quantum effects, such as
interference, become apparent on a macroscopic scale. The critical
temperature for a gas consisting of non-interacting relativistic
particles is given by \citep{bh96}:
\begin{equation}
T_c\sim (\frac{n_{ch}}{3 m})^{1/2},
\end{equation}
where the Boltzmann's constant and speed of light have been set to
unity.  Given the extremely low particle mass assumed here, $T_c$ is
derived from the relativitic Bose-Einstein particle-antiparticle
distribution with the chemical potential set to particle mass $m$. Here the "charge"
density $n_{ch}\equiv n_+ - n_-$, where $n_+$ and $n_-$ are the
number densities of particles and antiparticles in excited states.
On the other hand, we have $n_{ch}\sim (m/T) n_+$, and it follows
that $T_c\sim(\frac{n_+}{3T})^{1/2}$.  Note that $n_+$ scales as
$a^{-3}$ and $T$ as $a^{-1}$, and it follows $T_c$ scales as $a^{-1}$.
It means that when $T$ is below $T_c$ at some time after a phase
transition, the temperature will remain sub-critical in any later epoch.
As an estimate, if we assume one percent of ELBDM to be in the excited states after its decoupling, the current critical temperature becomes
\begin{equation}
T_c=3 \times 10^{-14}(\frac{m}{eV})^{-1/2} (\frac{T}{eV})^{-1/2} eV.
\end{equation}
Substituting $m\sim 10^{-22}$eV and $T\sim 10^{-4}$eV, the same as the present
photon temperature, we find that the current critical temperature
$T_c=0.3 eV >> T$.  Hence ELBDM, if exists and accounts for the dark matter,
may very well be in the BEC state ever since a phase transition in the early universe.
Despite ELBDM particles in the excited state are with a relativistic temperature, almost all particles are in the ground state and described by a single non-relativistic wave function.

\subsection{Basic Analysis}

The Lagrangian of non-relativistic scalar field in the comoving
frame is
\begin{equation}
L={a^3\over 2}[i\hbar(\psi^*{\partial\psi\over\partial t}-
\psi{\partial\psi^*\over\partial t}) +{\hbar^2\over
a^2m}(\nabla\psi)^2-2mV\psi^2],
\end{equation}
and the equation of motion for this Lagrangian gives a modified form
of Schr\"{o}dinger's Equation \citep{slopez03} :
\begin{equation}
i\hbar \frac{\partial\psi}{\partial t}=-\frac{\hbar ^{2}}{2
a^2m}\nabla ^{2}\psi + m V\psi,
\end{equation}
where $\psi\equiv \phi ({n_0}/a^3)^{-1/2}$ with $\phi$ being the
ordinary wave function, $n_0$ the present background number density
and $V$ is the self-gravitational potential obeying the Poisson
equation,
\begin{equation}
\nabla ^{2} V = 4 \pi G a^2\delta\rho = {4 \pi G \over a}{
\rho_0}(|\psi|^{2} -1).
\end{equation}
The only modification to the conventional Schr\"{o}dinger-Poisson
equation is the appearance $a^{-1}$ associated with the comoving
spatial gradient $\nabla$, and the probability density
$|\psi|^2$ to be normalized to the background proper density $\rho /m$.
In the above,
\begin{equation}
\rho_{0}\equiv \frac {3 H_{0} ^2}{8 \pi G} \Omega_{m}=m n_0
\end{equation}
is the background mass density of the universe.

To explore the nature of the ELBDM, we first adopt the
hydrodynamical description to investigate its linear evolution.  This
approach is not only more intuitive than the wave function
description, its advantage will also become apparent later.
Let the wave function be
\begin{equation}
\psi = \sqrt{\frac{{n}}{n_0}} e ^{i \frac{S}{\hbar}},
\end{equation}
where ${n}=\bar{n} a^3$, the comoving number density and ${\bar{n}}$ is the proper number density. The quadrature of Schr\"{o}dinger's equation can
be split into real and imaginary parts, which become the equations
of accerleration and density separately,
\begin{equation}
\frac{\partial}{\partial t} {\bf v}+ \frac{1}{a^2}{\bf
v}\cdot\nabla{\bf v} + \frac{\nabla V}{m} - \frac {\hbar ^{2}}{2 m^2
a^2} \nabla (\frac {\nabla ^{2} \sqrt{{n}}}{\sqrt{{n}}})=0
\end{equation}
\begin{equation}
\frac{\partial {n}}{\partial t} +{1\over a^2}\nabla \cdot
({n} {\bf v})=0,
\end{equation}
where ${\bf v}\equiv\nabla S/m$ is the fluid velocity. There is a
new term depending on the third-order spatial derivative of the wave
amplitude $\sqrt{n}$ in the otherwise cold-fluid force equation.
This term results from the "quantum stress" that acts against
gravity, and it can be cast into a stress tensor in the energy and
momentum conservation equation \citep{chiu98,chiu00}. The quantum stress
becomes effective only when the spatial gradient of the structure is
sufficiently large.

The fluid equations, Eqs.(5),(8) and (9), are linearized and combined
to yield
\begin{equation}
\frac{\partial}{\partial t}a^2\frac{\partial}{\partial t}\delta
n-\frac{3 {H_0}^2 \Omega_m}{2a}\delta n
+\frac{\hbar^2}{4m^2a^2}\nabla^2\nabla^2\delta n=0.
\end{equation}
Upon spatially Fourier transforming $\delta n$, it follows
\begin{equation}
\frac{d}{d t}a^2\frac{d n_k}{d t}-({3 {H_0}^2 \Omega_m \over {2a}})
n_k +\frac{\hbar^2k^4}{4m^2 a^2}n_k=0,
\end{equation}
which can be recast into
\begin{equation}
x^2\frac{d^2}{dx^2}n_k+(x^2-6)n_k=0,
\end{equation}
and the solution to this equation is
\begin{equation}
n_k =\frac{(3\cos{x}-x^2\cos{x}+3x\sin{x})}{x^2}
\end{equation}
where $x\equiv \hbar k^2/(m\sqrt{ {H_0}^2 a})$ and $a=(t/t_0)^{2/3}$
and $\Omega_m = 1$ appropriate for early universe have been assumed.
In the small-$k$
limit, $x$ is small and $n_k\sim x^{-2}$, which grows in time as
$a$; for large $x$ the solution oscillates. Fig.(1) depicts the
solution, Eq.(13).   From Eq.(12) we can easily identify the oscillating
solutions when $x^2 \geq 6$, thereby defining the
Jeans wave number:
\begin{equation}
k_{J} = (6a)^{1/4}(\frac{mH_{0}}{\hbar})^{1/2}.
\end{equation}

Beyond the Jeans wave number, the perturbation is suppressed by
quantum stress. Moreover, the Jeans wave number scales as $a^{1/4}$
and is proportional to $m^{1/2}$ (Hu, Barkana \& Gruzinov 2000). We shall come back in a later section to examine up to what evolutionary stage the linear solution of $n_k$ can remain valid.

Next, we linearize the Schr\"{o}dinger-Poisson equation to derive
a governing equation for an alternative wave-function representation.
The wave function can be separated into real and imaginary parts,
\begin{equation}
\psi = 1+ R+ iI,
\end{equation}
with $R$, $I \ll 1$. In the linear regime, we have the real part of
linearized Schr\"{o}dinger's equation
\begin{equation}
-\hbar \frac{ \partial}{\partial t}I =-\frac{\hbar^2}{2 a^2
m}\nabla^{2}R + m V,
\end{equation}
and the imaginary part
\begin{equation}
\hbar \frac{\partial R}{\partial t} = -\frac{\hbar^2}{2 a^2
m}\nabla^{2}I.
\end{equation}
The Poisson Equation becomes
\begin{equation}
\nabla^2 V= \frac{8\pi G}{a} \rho_0 R.
\end{equation}
The spatial Fourier components of gravitational potential and
$\psi$ satisfy
\begin{equation}
V_k=-\frac{8\pi G \rho_0}{a}\frac{R_k}{k^2},
\end{equation}
\begin{equation}
-\hbar \frac{d}{dt}I_k =\frac{\hbar^2 k^{2}}{2 a^2 m}R_k + m V_k,
\end{equation}
and
\begin{equation}
I_k=\frac{2 a^2 m}{\hbar k^2} \frac{dR_k}{dt}.
\end{equation}
Combing the above, we have, as Eq. (11), that
\begin{equation}
\frac{d}{d t} a^2 \frac{d}{d t}R_k
 - (\frac{3 {H_0}^2 \Omega_m}{2a})R_k + (\frac{\hbar^2 k^4}{4 m^2 a^2} R_k=0,
\end{equation}
and $R_k$ has, up to a constant factor, the same solution as $n_k$.
Note that since $dR_k/dt = \dot{a} R_k/a$ for low-$k$ modes, it follows
that $|I_k|= (2mH_0 a^{1/2}/\hbar
k^2)|R_k|=({k_{J}}^2/\sqrt{3}k^2)|R_k| \gg |R_k|$. This feature will
serve as one of the indicators for the validity of the linear regime
in the wave function representation.

\begin{figure}
 \begin{center}
  \includegraphics[width=12cm,angle=0]{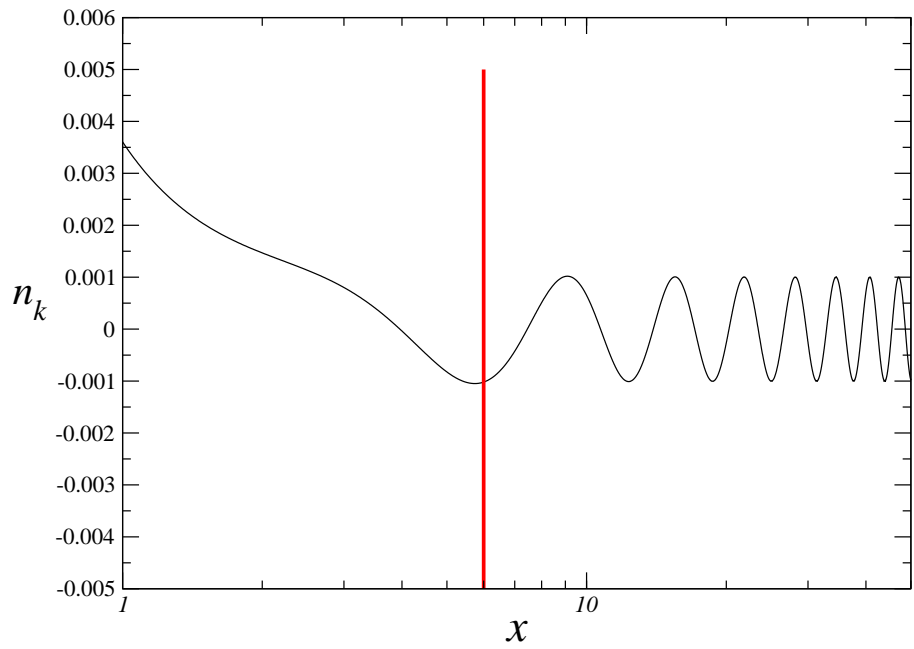}\\
  \caption{The solution of the linear perturbation given by Eq.(6). The vertical
line labels the location of squared scaled Jean's wavenumber at $x=6$.}\label{fig:Solution}
 \end{center}
\end{figure}

\begin{figure}
 \begin{center}
  \leavevmode
  \includegraphics[width=12cm,angle=0]{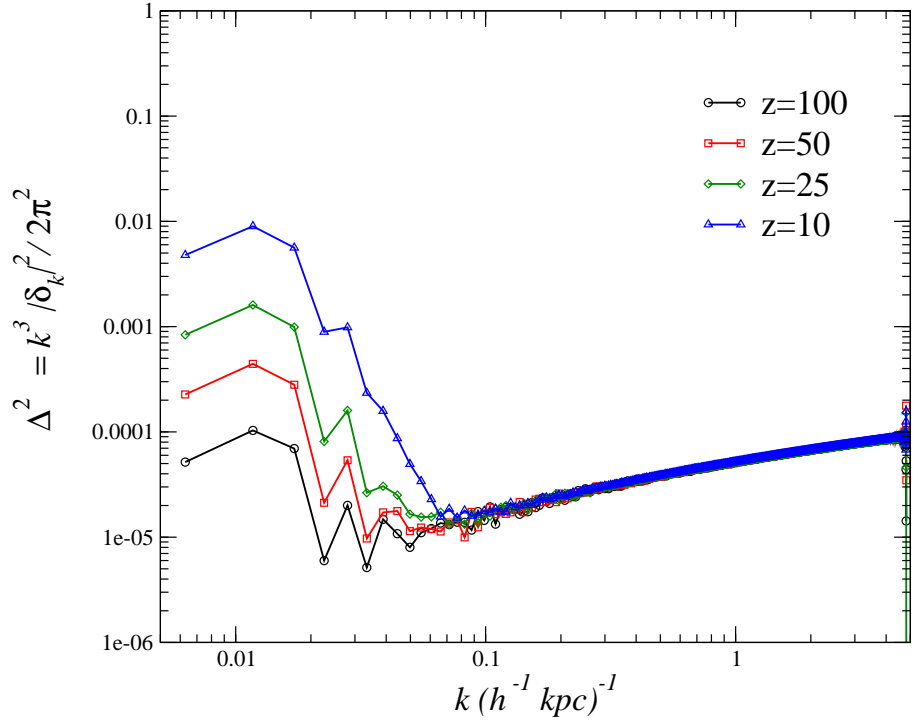}\\
  \caption{The linear evolution of the power spectrum from $z=100$ to $z=10$ in an 1 $h^{-1}$Mpc box. The low-$k$ power obeys the linear scaling $\propto a^2$. }\label{fig:Power_Evolve_Linea}
 \end{center}
\end{figure}

\section{Numerical Scheme and Simulations}

\subsection{Numerical Scheme}

We normalize the length to the computational grid size, $\Delta$, and
further define $\eta=(m \Delta^2 H_{0})/{\hbar}$. The value of
$\eta$ determines the size of Jean's length relative to the computational grid size.  Set $\nabla =
\frac{1}{\Delta} \widetilde{\nabla}$. The dimensionless
Schr\"{o}dinger-Poisson equations becomes
\begin{equation}
i\frac{\partial\psi}{\partial \tau}= -\frac{1}{2 a^2
\eta}\widetilde{\nabla} ^{2}\psi + \frac{3\Omega_m \eta}{2a} U\psi,
\end{equation}
and
\begin{equation}
\widetilde{\nabla}^{2} U = (\vert \psi \vert ^{2}- 1),
\end{equation}
where $U(x)={V(x)}/(\frac{3\Omega_{m} \eta}{2a})$, the
dimensionless gravitational potential, and $\tau=H_{0} t$.

Given a Hamiltonian $\bf {H}$, one can evolve the wave function
through (23). It is simply a unitary transformation of the system,
\begin{equation}
\psi^{j+1}= e^{-i{\bf{H}}dt}\psi^{j}.
\end{equation}
We use the \emph{pseudo-spectral method} to solve the Schr\"{o}dinger
equation in the comoving box. Let ${\bf{K}}$ be the kinetic energy
operator (${\bf{K}}=-\frac{1}{2\eta a^2}\widetilde{\nabla}^2
\rightarrow {{k}}^2/2\eta a^2$ in Fourier space)
and $\bf{W}$ the potential operator(${\bf{W}}=\frac{3\Omega_m}{2
a}U$ in real space). The evolution is then split into

\begin{equation}
e^{-i{\bf{H}}dt}=e^{-i({\bf{K}+\bf{W}})dt}=1-i({\bf{K+W}})dt-\frac{1}{2}{\bf(K^2+KW+WK+W^2)}
dt^2+O(dt^3).
\end{equation}
On the other hand, we need to consider the non-commutative relation
between ${\bf{K}}$ and ${\bf{W}}$, where

\begin{equation}
e^{-i{\bf{K}}{dt}}e^{-i{\bf{W}} dt}=1-i({\bf{K+W}})
dt-\frac{1}{2}{\bf{K}}^2 dt^2-\frac{1}{2}{\bf{W}}^2 dt^2-{\bf{KW}}
dt^2+O(dt^3),
\end{equation}
\begin{equation}
e^{-i{\bf{W}}{dt}}e^{-i{\bf{K}} dt}=1-i({\bf{W+K}})
dt-\frac{1}{2}{\bf{W}}^2 dt^2-\frac{1}{2}{\bf{K}}^2 dt^2-{\bf{WK}}
dt^2+O(dt^3).
\end{equation}
It follows that we obtain, to the second order accuracy,

\begin{equation}
e^{-i{\bf{(K+W)}}dt}\approx \frac{1}{2}[e^{-i{\bf{K}}
dt}e^{-i{\bf{W}} dt}+e^{-i{\bf{W}} dt}e^{-i{\bf{K}} dt}],
\end{equation}
which will be adopted to advance the time steps.

For each time step, the kinetic energy
operator is calculated in the Fourier domain,
\begin{equation}
{\psi_k}^{j+1}=e^{-i\frac{{k}^2}{2\eta
a^2}dt}{\psi_k}^{j}
\end{equation}
and $\psi$ is advanced in real space with the
potential energy operator,
\begin{equation}
{\psi({\bf x})}^{j+1}=e^{-i\frac{3 \Omega_m \eta U}{2a}dt}{\psi({\bf
x})}^{j}.
\end{equation}
To ensure numerical stability, we restrict the magnitude of $dt$
that rotates the phase angle of wave function less than
$\frac{\pi}{4}$ in each time step,
\begin{equation}
dt \leq \frac{\pi}{2} \frac{\eta a^2}{{k_{max}}^2},
\end{equation}
\begin{equation}
dt \leq |\frac{\pi a }{6 \Omega_m \eta U_{max}}|.
\end{equation}
In the early stage($a \sim 10^{-3}$), the stability condition is
governed by the kinetic energy term, where
${k_{max}}^2=3{\pi}^2$ and $dt \leq {{(6 \pi)}^{-1}} (\eta
a^2)$. At the late time, the gravitational potential becomes ever
deeper, and therefore $dt$ is controlled by the potential energy,
where $U_{max}$ is the greatest value of potential in the real
space.

\subsection{Simulation Scale}

We prepare the initial conditions with CMBFAST \citep{cmbfast96} at $z=1000$ with $\Lambda$CDM cosmology.  Such initial conditions differ from that of \citet{hu00},
where the Compton length of ELBDM already has imprints on the power spectrum
at $z=1000$.  We choose this initial condition because only a few low-$k$ modes can grow for our choice of Jean's length and the details of initial power spectrum are irrelevant at the late time.
The simulations run up to $1024^{3}$-grid resolution in a 1 $h^{-1}$Mpc comoving box.  For simulations in a much larger box, the background density averaged over an 1$h^{-1}$ Mpc box can often change with time, the so-called environment effects, where galaxies prefer to form in regions of high background density.  Here, we ignore the environment effect by fixing $\Omega_m=0.3$.
We let the dimensionless parameter $\eta=1.22 \times 10^{-2}$ and $4.88\times 10^{-2}$ for the $1024^3$ and $512^3$ simulation boxes, which give a Jeans wavelength 50 kpc at $z=0$.  This value of $\eta$ corresponds to $m \sim 2.5 \times 10^{-22}eV$.  In the rest of this paper, we shall report only the simulation results of the highest resolution.

\section{Results}

\subsection{Validity of Linear Perturbation Theory}

We depict the early evolution of density power spectrum in Fig (2)
from $z=100$ to $z=10$. The density power spectrum increases as $a^2$
for modes with wave number $\ll k_{J}$, and $k_{J}$ indeed increases
as $a^{1/4}$. These features are in agreement with what the linear
perturbation theory predicts.

On the other hand, we show the early evolution of $4|R_k|^2$ and
$4|I_k|^2$ in Fig.(3) from $z=400$ to $z=10$. As expected the
low-$k$ modes grow initially, while the high-$k$ modes are
suppressed by quantum pressure. It is surprising to find that the
linear evolution of $\psi_k$ is valid only for a short period of
time before $z = 200$. After that, the wave function deviates from
what the linear theory predicts. In particular, the linear theory of
wave function representation predicts that $R_k$ and $I_k$ grow as
$a$ and $a^{3/2}$ respectively, and $|I_k|^2=({k_{J}}^4 /3 k^4)
|R_k|^2 \gg |R_k|^2$ for low-$k$ modes.  This prediction agrees with
the simulation result only when $z>200$.
At a somewhat later epoch
than $z=200$, we observe that the difference between $|I_k|^2$ and $|R_k|^2$
diminishes, and at $z=10$ we find $|I_k|^2 \approx |R_k|^2$.

This problem is also manifested in the growth rate of the linear
density power spectrum $|n_k|^2\approx 4|R_k|^2$. It is found that
$4|R_k|^2$ indeed scales as $a^2$ before $z=200$. When $z < 200$,
the quantity $4|R_k|^2$ grows much faster than $a^2$, and $4|R_k|^2$
becomes about one order of magnitude greater than what the linear
theory predicts at $z=100$.  That is, the density power spectrum $|n_k|^2$
vastly deviates from, and is much less than, $4|R_k|^2$ even since early on in the
evolution.

To examine this peculiar feature, we construct the imaginary part of wave function $I(\bf{x})$ from $I_k$ of a few
lowest $k$ modes and depict $|(I^2)_k|^2$ on the same plot as
$4|R_k|^2$ at $z=100$ in Fig.(3). It is found that
$|{(I^2)}_k|^2$ coincides with $4|R_k|^2$ at low-$k$. Since $n_k \approx 2R_k + (I^2)_k$, it follows that $(I^2)_k$ has the opposite
sign but approximately the same magnitude as $2R_k$ so that
the two terms of $n_k$ almost cancel to yield $|n_k|^2 \ll 4|R_k|^2$.
Thus nonlinearity already sets in as early as $z=100$ in
the wave function representation.  On the other hand, the fluid representation
does not have such a problem. We find that $|n_k|^2$ of low-$k$ modes in the
fluid representation agrees with what the linear fluid theory predicts even as late as
$z = 1$, despite that the high-$k$ modes already become nonlinear.
The difference arises from that $(\nabla S/m)(\equiv \bf{v})$ in the
fluid representation remains small at low-$k$, in contrast to $I=S$
in the wave function representation which has a large amplitude for low-$k$ modes
even at high $z$.
\begin{figure}
 \begin{center}
  \leavevmode
  \includegraphics[width=12cm,angle=0]{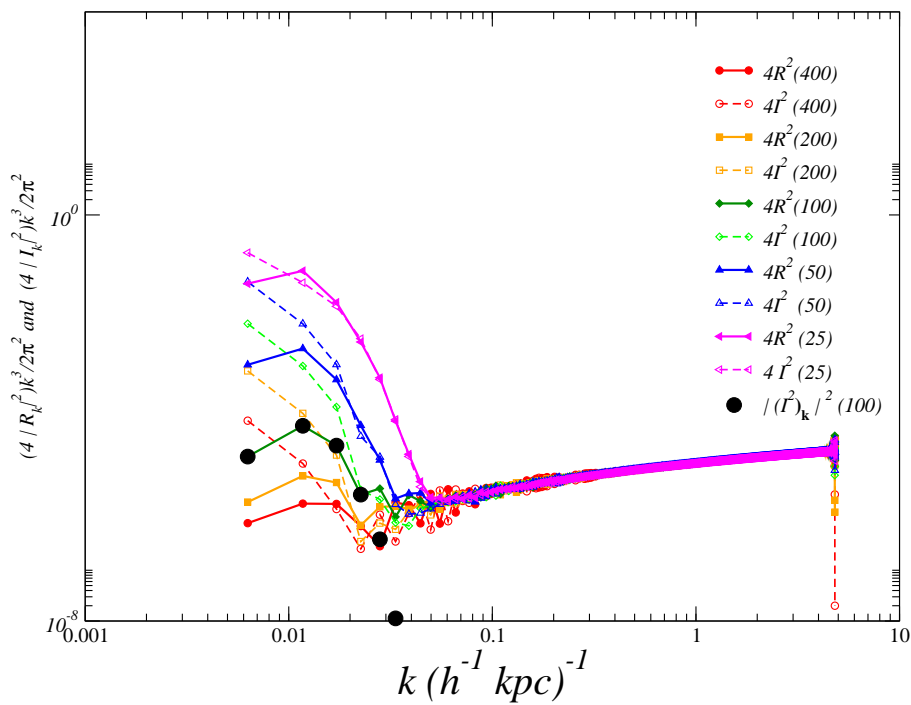}\\
  \caption{Evolution of $4|R_k|^{2}(z)$ and $4|I_k|^{2}(z)$ to check against the linear theory. Deviation from the linear theory is evident from z=200 on. Black dots are $|(I^2)_k|^2$ at $z=100$ constructed from few low-$k$ modes to show the cancellation between $2R_k$ and $(I^2)_k$ so as to make $n_k \ll 2R_k$. }\label{fig:4R2_4I2}
 \end{center}
\end{figure}

\subsection{Weakly Nonlinearity Regime}

\begin{figure}
\begin{center}
  \includegraphics[width=11cm,angle=0]{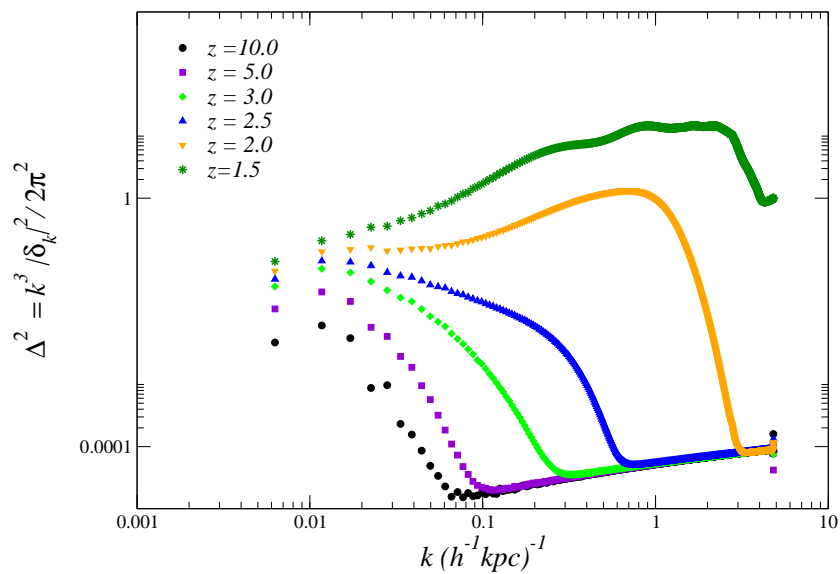}
  \caption{The weakly nonlinear evolution of the power spectrum from
$z=10$ to $z=1.5$, where the high-$k$ modes are seen to be nonlinearly excited.
}\label{fig:z_10_z_1}
\end{center}
\end{figure}

 Shown in Fig.(4) is the evolution of $|n_k|^2(=|2R_k+(R^2+I^2)_k|^2)$ for $1.5<z<10$.
The initial $n_k$ of high-$k$ mode has been linearly suppressed and is later replaced by the high-$k$ modes that are nonlinearly generated beginning around $z=5$. The nonlinear coupling arises from the coupling $V\psi$ in the Schr\"{o}dinger equation. Since $V$ is dominated by low-$k$ modes, the nonlinear coupling transfers modal energy local in $k$ space from one mode to the neighboring mode, and from low $k$ to high $k$. The gravitational potential $V$ is barely evolving in the weakly nonlinear regime, and hence the dynamics in this regime is for the wave function to settle into almost static potential wells.

\begin{figure}
 \begin{center}
  \leavevmode
  \includegraphics[width=8cm]{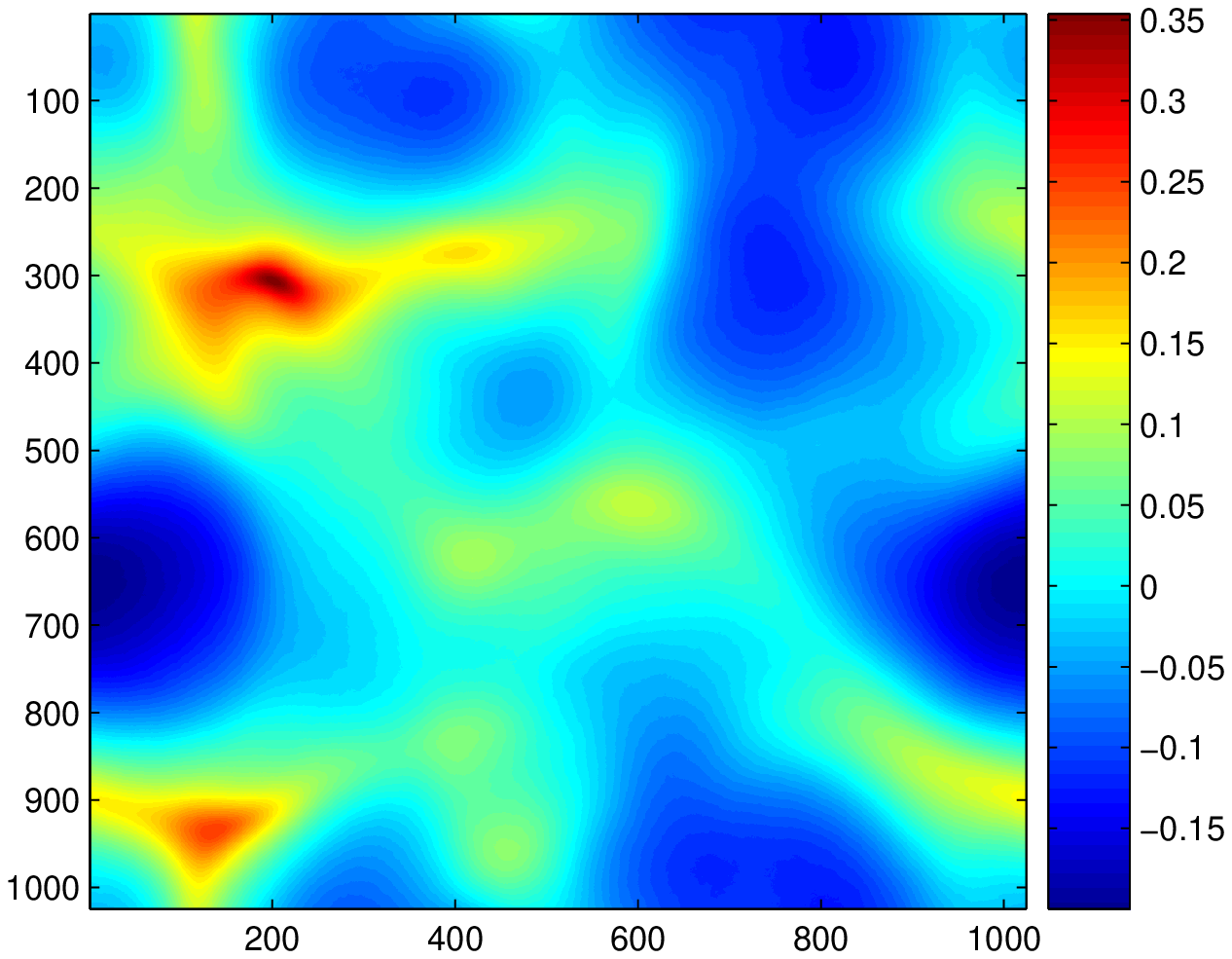}
  \includegraphics[width=7.8cm]{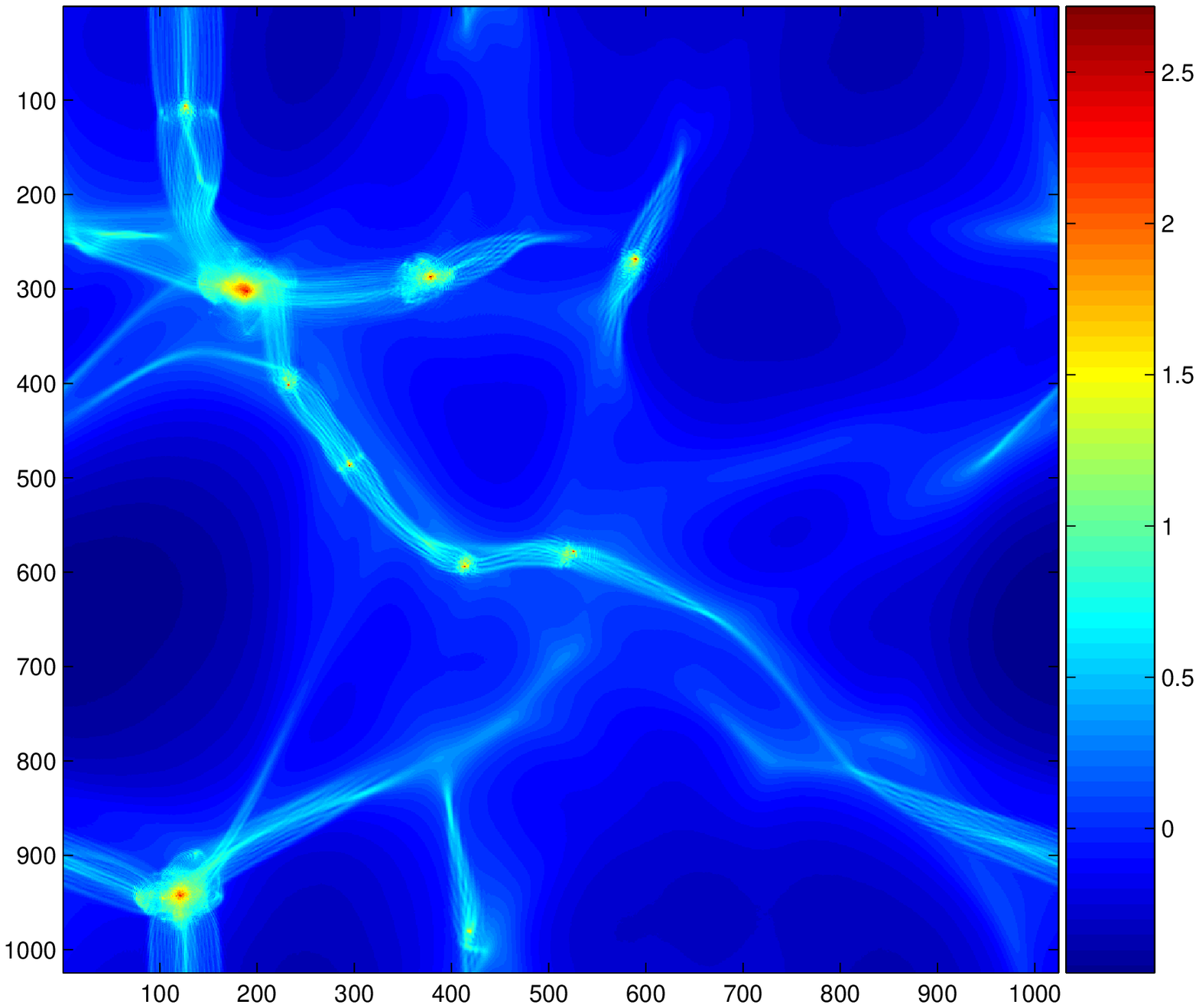}\\
  \caption{Two-dimensional projections of density in real space in a 1$h^{-1}$Mpc comoving box at z=3 (left panel) and z=0 (right panel). Halo A and B are at the top left and bottom left.}
  \label{fig:2D_project1}
 \end{center}
\end{figure}

We note that if $V$ were exactly static, the Schr\"{o}dinger equation would have been linear and the coupling from low to high $k$ modes would simply be the linear evolution of wave function starting with a non-eigenstate.  This argument may explain why the linear theory
of low-$k$ modes works so well even after high-$k$ modes are excited in this regime. Shown in the left panel of Fig.(5) is the real-space configuration of $\delta n$ at $z=3$ in this weakly nonlinear regime, where the wave function settling into individual quasi-static potentials well is underway.

\subsection{Strong Nonlinearity Regime}

Plotted in Fig.(6) is the evolution of density power spectrum after $z=1.5$. This is the strongly nonlinear regime where the gravitational potential develops deep wells at the collapsed cores.
To illustrate the contribution of the few collapsed objects to the final high-$k$ power spectrum, $P(k,0)$, at $z=0$, we remove all matter outside the virial radii of these collapsed halos and construct the power spectrum of these artificial objects, ${P_h}(k,0)$. The power spectrum of the removed matter, $P_{b}(k,0)$, is also constructed for reference. The two power spectra along with the original power spectrum are depicted in Fig.(7) for comparison. Clearly the bound objects contribute to almost all the power contained in the original power spectrum, except for the low-$k$ modes that are contributed dominantly by $P_b(k,0)$.  These few lowest-$k$ modes are grown out of the initial noise and remain so in the final configuration. That is, despite that the initial condition possesses many independent degrees of freedom, the final configuration has only a few degrees of freedom, where almost all randomly placed, small collapsed objects seen in standard CDM simulations are entirely suppressed.

During the final collapse phase, the core undergoes large-scale oscillations that send out waves to remove excess angular momentum deposited into the core region, rendering the core to settle into an almost stationary configuration in the physical space.  Fig.(8) shows the wavy structures of this nature around the collapsed halo A at z=1.  Even in the quasi-stationary state of these halos at $z = 0$, we find that this wave phenomenon is still pronounced around the halos, as will be discussed later.

There are two collapsed halos, A and B, of mass $5.7 \times 10^{9} M_{\bigodot}$
and $5 \times 10^{9} M_{\bigodot}$ respectively, at $z=0$ in our simulation as shown in the right panel of Fig.(5). Halo B is subject to major merger around $z=0.7$. Shown in Fig.(9) are halo B before and after the major merger. The final density profiles of halos A and B are plotted in Fig.(10).  Interestingly, they both develop singular cores, in spite of the presence of quantum pressure. Both power-law singular cores have a power index $-1.4$, reminiscent of that of the standard cold dark matter \citep{ginamor00}. The density profiles of the outskirts also obey power law, with a power index $-2.5$, slightly shallower than that produced by the cold dark matter \citep{nfw97}. Similar density profiles arising from different formation processes, i.e., accretion versus merger, suggest that the profile can be universal.

\begin{figure}
 \begin{center}
  \includegraphics[width=12cm,angle=0]{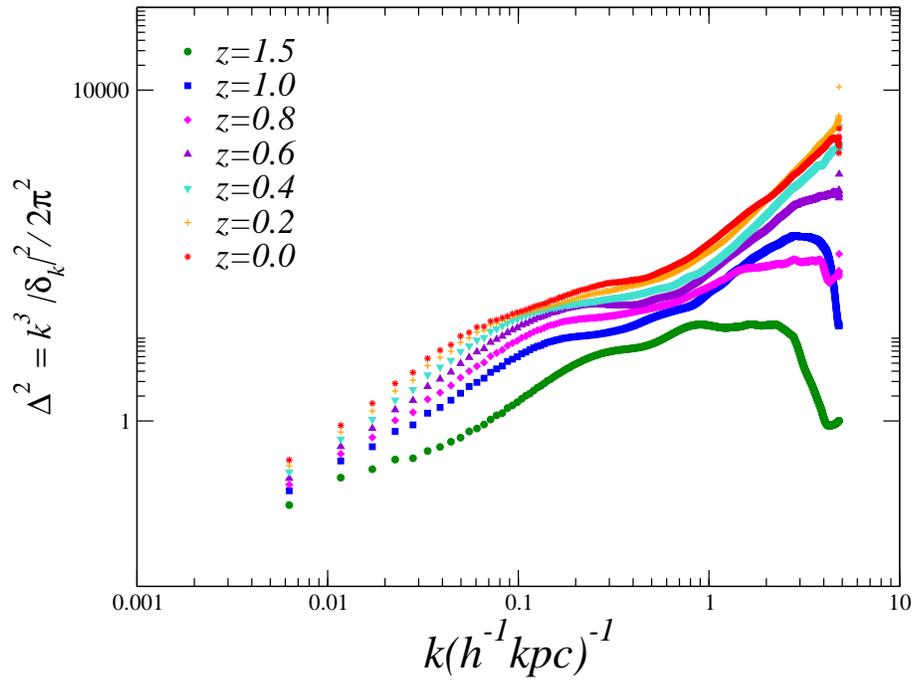}\\
  \caption{The nonlinear evolution of the power spectrum from
$z=1.5$ to $z=0.0$. The highest-$k$ modes acquire their full power after $z=0.4$, indicative of the creation of singular halo cores.
}\label{fig:z_1_z_0}
 \end{center}
\end{figure}

\begin{figure}
 \begin{center}
  \includegraphics[width=11cm,angle=0]{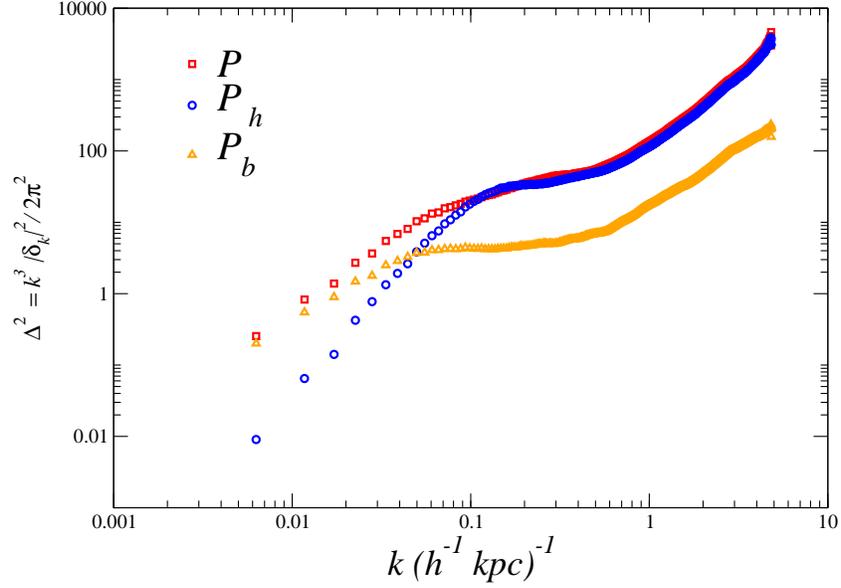}
  \caption{The comparison of the halo power spectrum $P_{h}$ (circle), the background power spectrum $P_{b}$ (triangle) and the full power spectrum $P$ (square) at $z=0$. Note that $P_{h}$ matches $P$ at high $k$  and $P_b$ matches P at low $k$. }\label{fig:Power_cut_vs_no_cut}
 \end{center}
\end{figure}

\begin{figure}
 \begin{center}
  \includegraphics[width=8cm,angle=0]{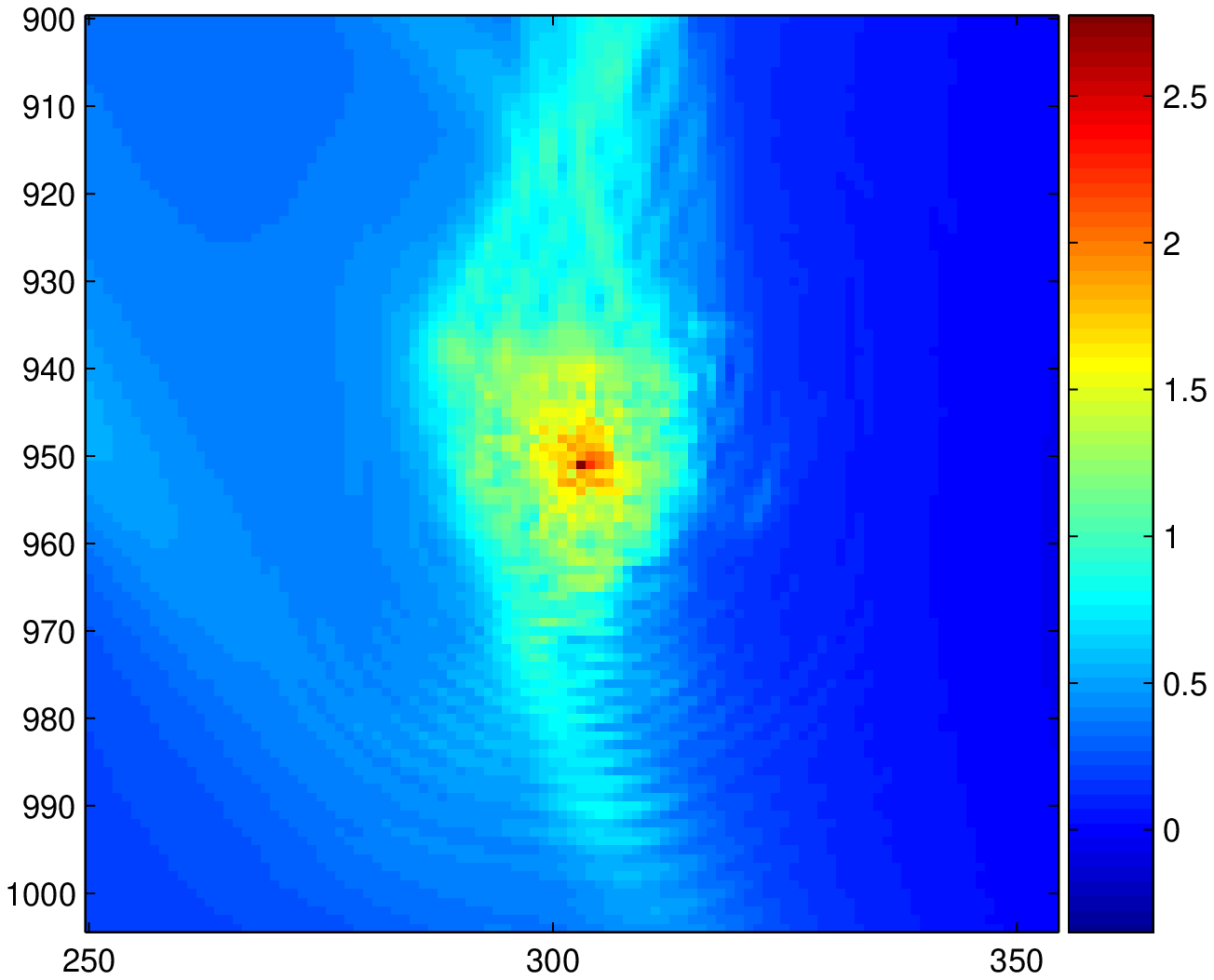}
  \includegraphics[width=8cm,angle=0]{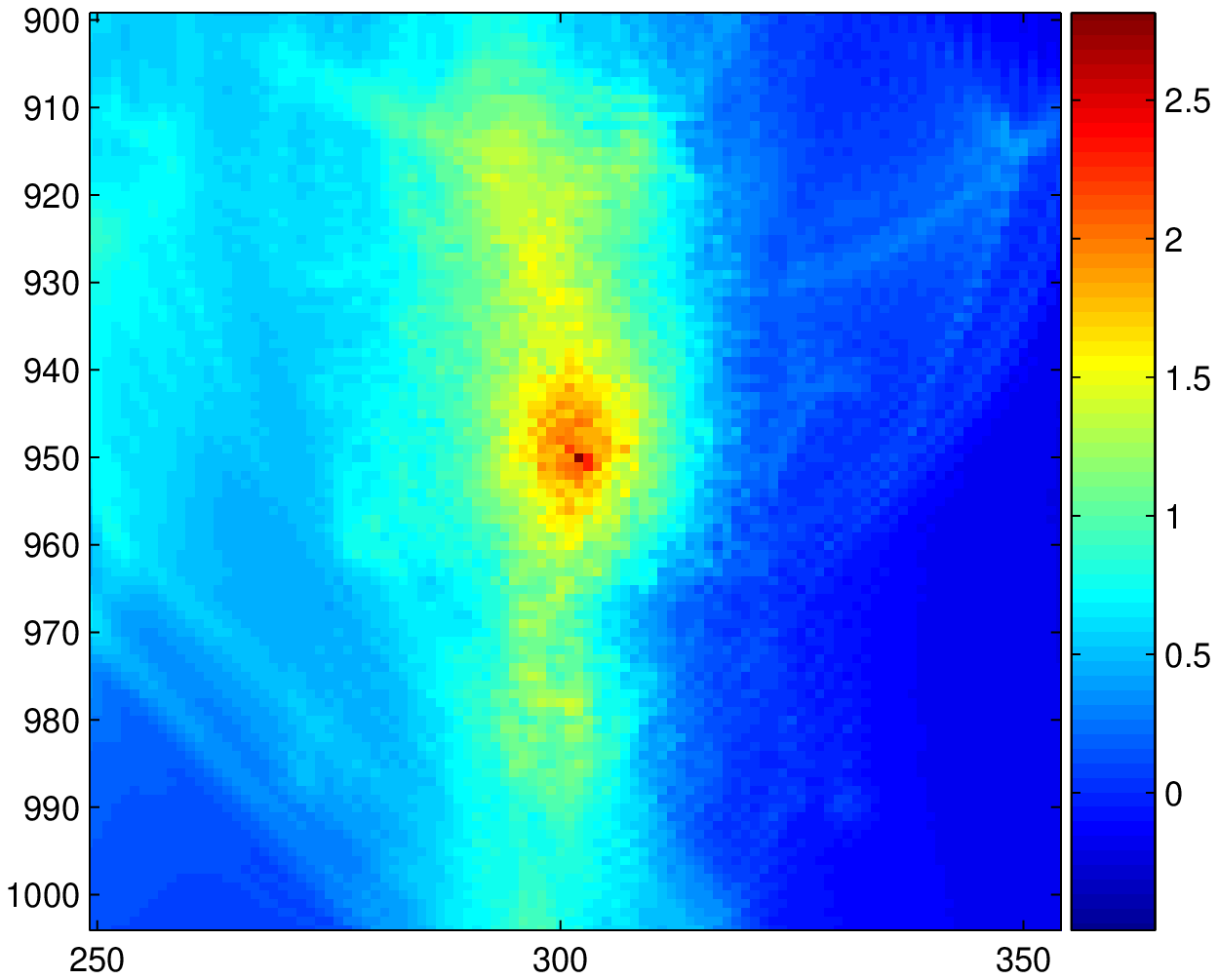}
  \caption{Waves are sent out from the collapsing halo A at $z=1$ (left panel); it develops an oblate singular halo at $z=0$ (right panel).  }\label{fig:wave_structure}
 \end{center}
\end{figure}

\begin{figure}
 \begin{center}
  \includegraphics[width=8cm,angle=0]{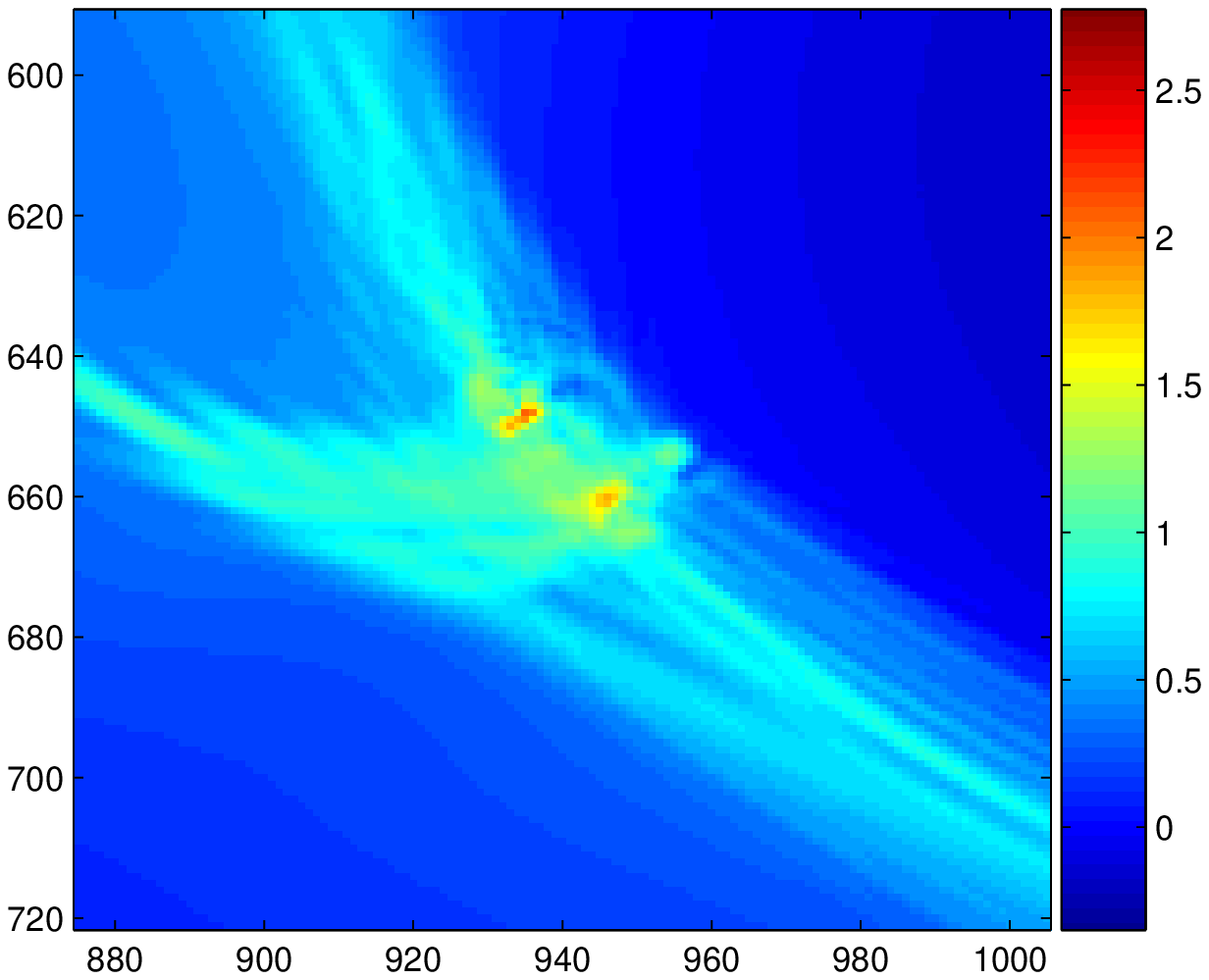}
  \includegraphics[width=8cm,angle=0]{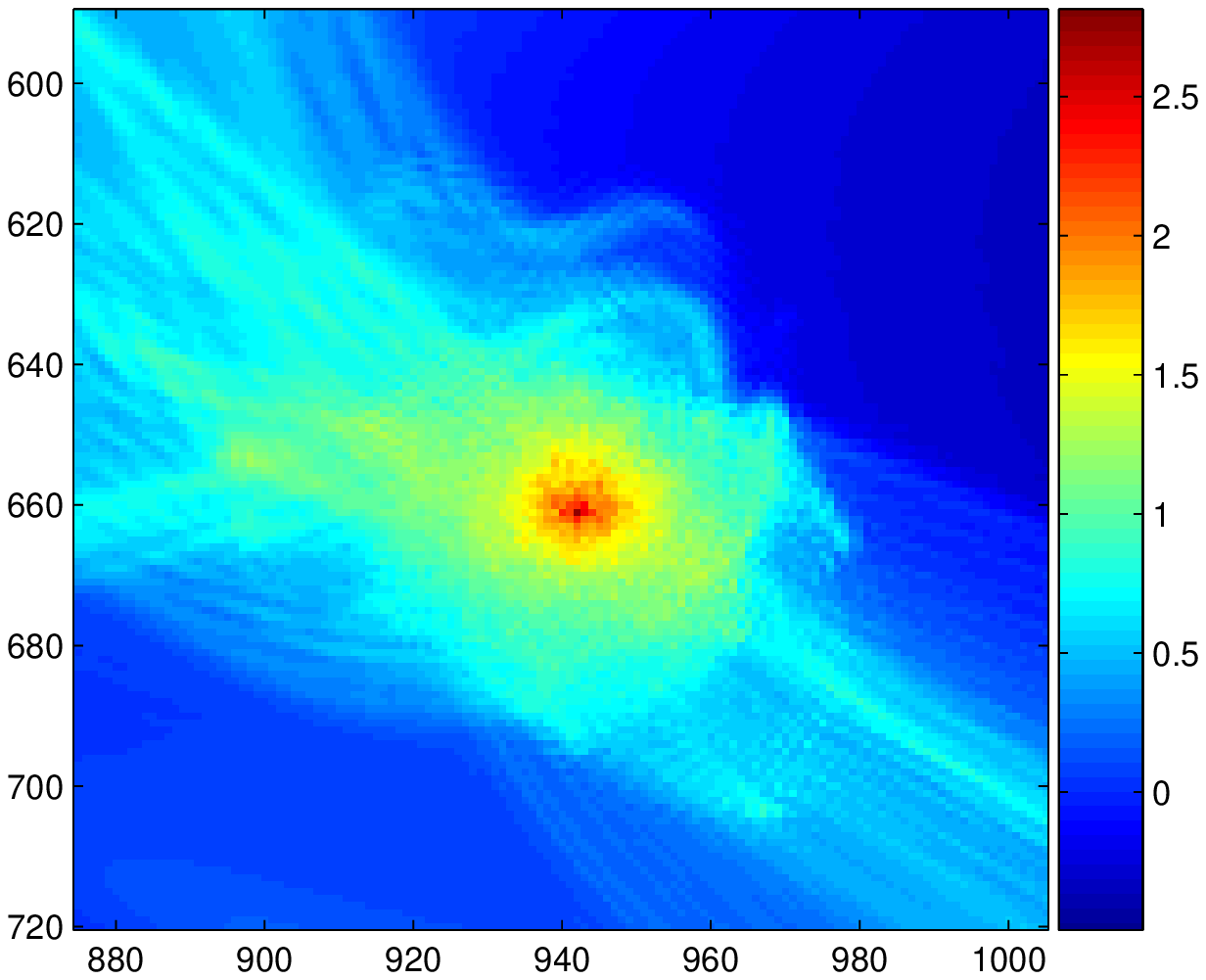}
  \caption{Halo B is subject to major merger at $z=0.7$. The left panel reveals the progenitors at $z=1$ and the right panel shows a singular halo with high degree of spherical symmetry at $z=0$.  }\label{fig:major_merger_before}
 \end{center}
\end{figure}

\begin{figure}
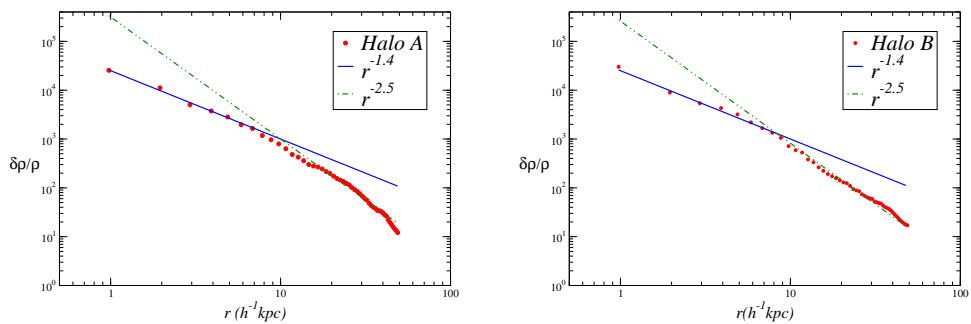

 \begin{center}
  \leavevmode
  \includegraphics[width=6cm]{f10a.eps}
  \hspace{0.5cm}
  \includegraphics[width=6cm]{f10b.eps}\hspace*{\fill}\\
  \caption{Density profiles of two massive halos at z=0. The left panel plots the profile of halo A and the right panel the profile of halo B. In both panels dot-dash lines and solid lines denote the power law of indices $-1.4$ and $-2.5$, respectively.}
  \label{fig:Density_Profile}
 \end{center}
\end{figure}

\begin{figure}
 \begin{center}
 \leavevmode
  \includegraphics[width=14cm,angle=0]{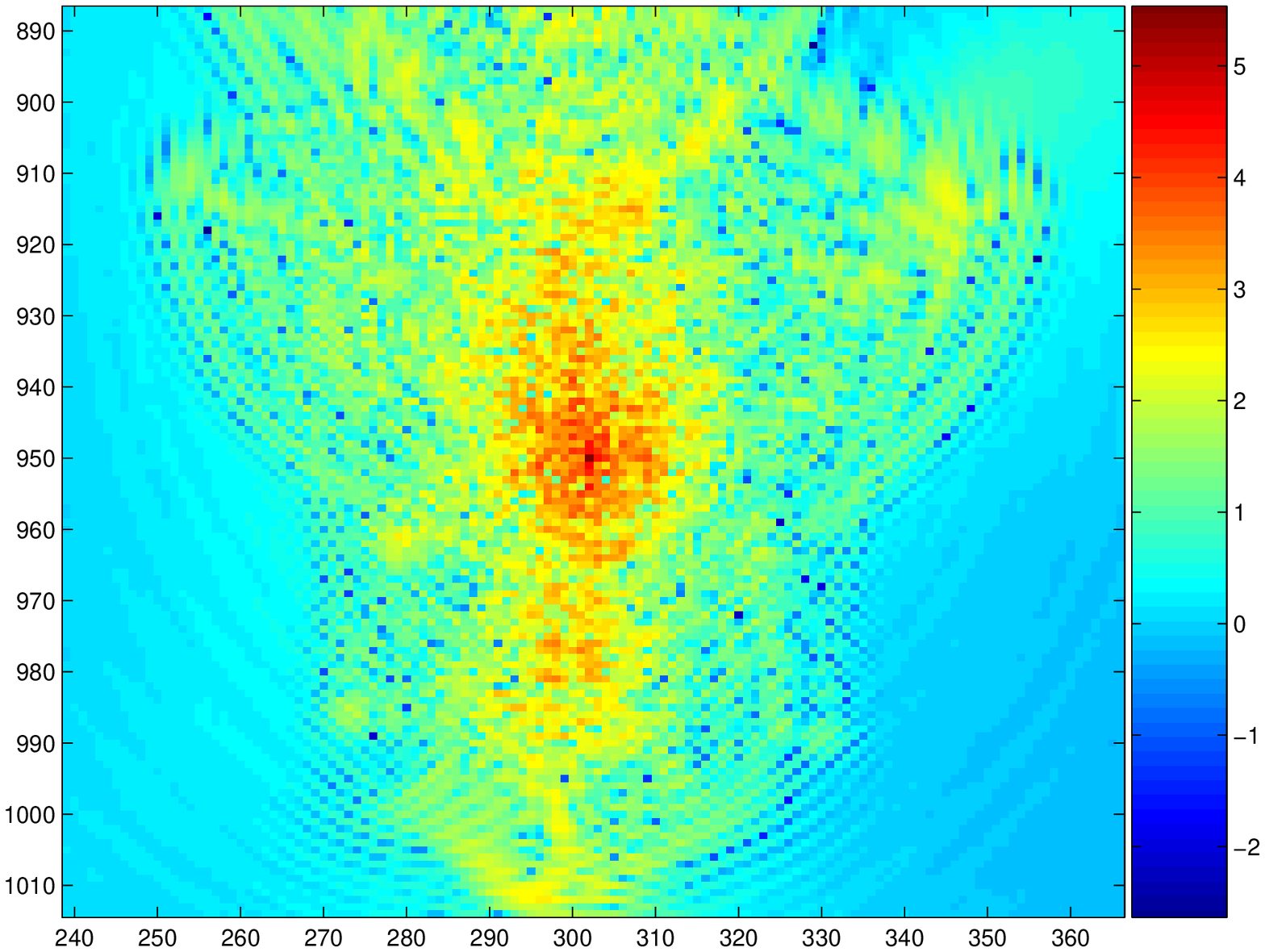}\\
  \caption{A two-dimensional slice of density for halo A through the core.}\label{fig:one_slice_image_A}
 \end{center}
\end{figure}

\begin{figure}
 \begin{center}
 \leavevmode
  \includegraphics[width=13cm,angle=-90]{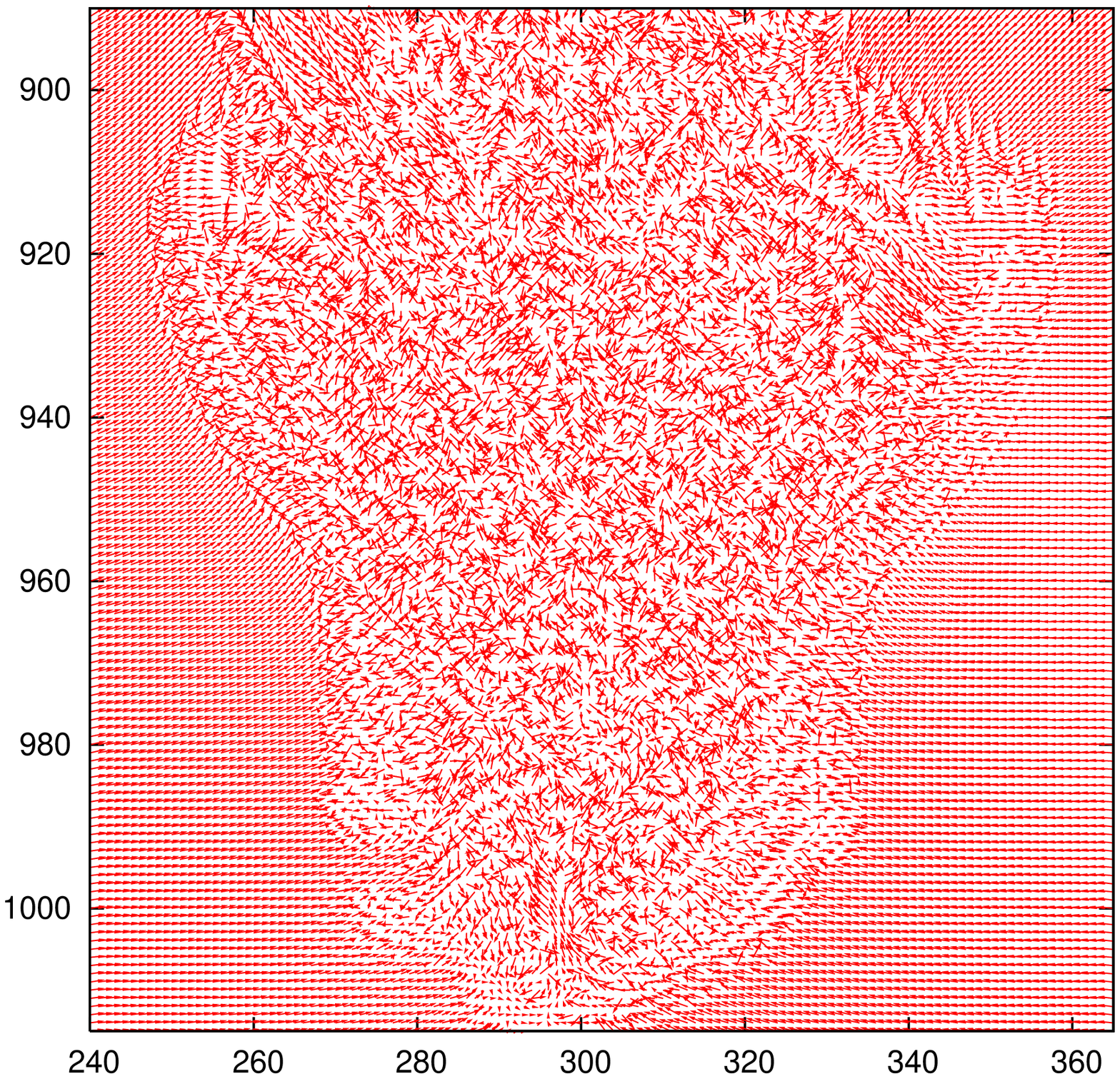}\\
  \caption{The same two-dimensional slice of the velocity field for halo A in the comoving frame.}\label{fig:Velocity_Field_Halo_A}
 \end{center}
\end{figure}

\begin{figure}
 \begin{center}
 \leavevmode
  \includegraphics[width=14cm,angle=0]{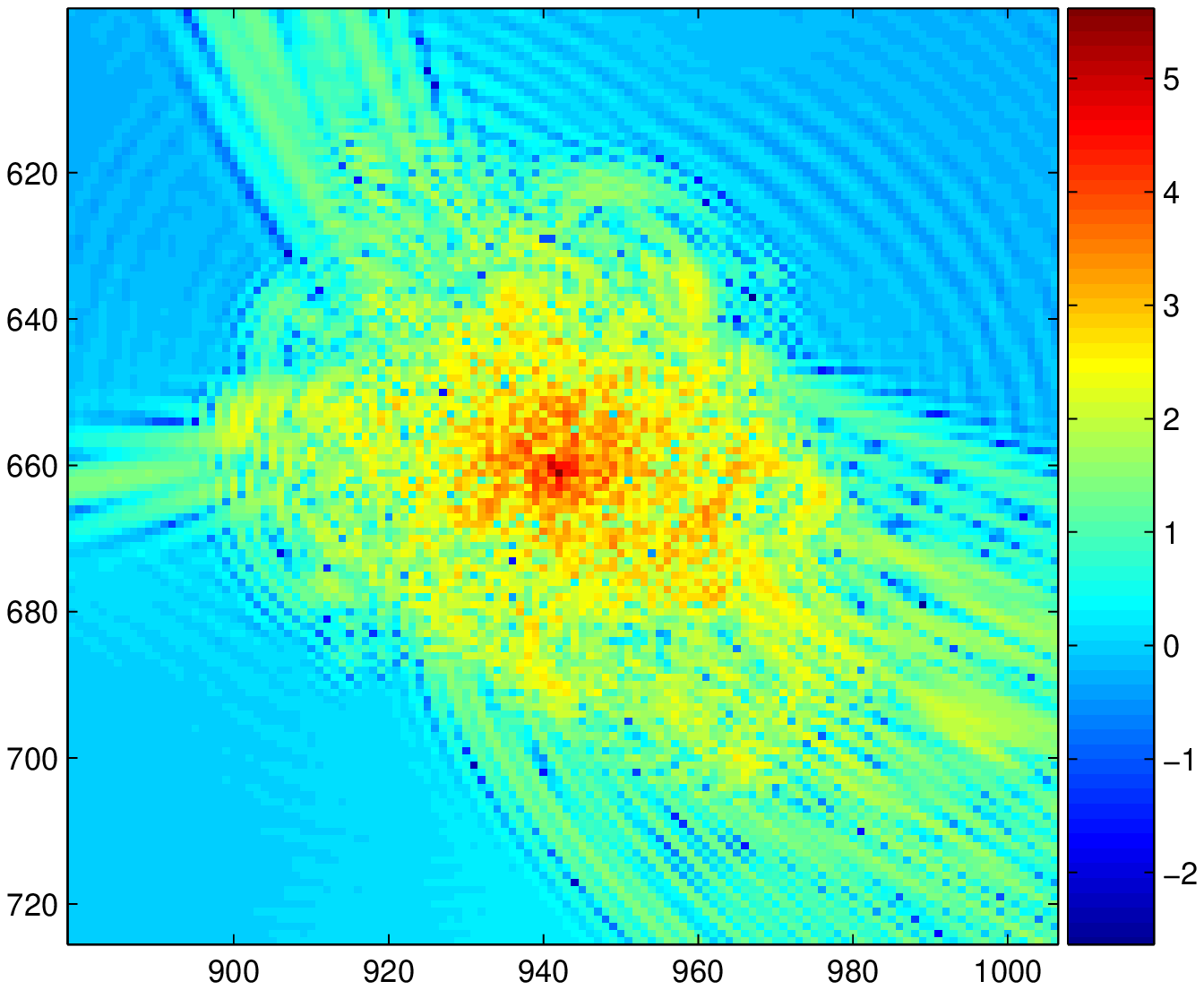}\\
  \caption{A two-dimensional slice of density for halo B through the core}\label{fig:one_slice_image_B}
 \end{center}
\end{figure}

\begin{figure}
 \begin{center}
 \leavevmode
  \includegraphics[width=13cm,angle=-90]{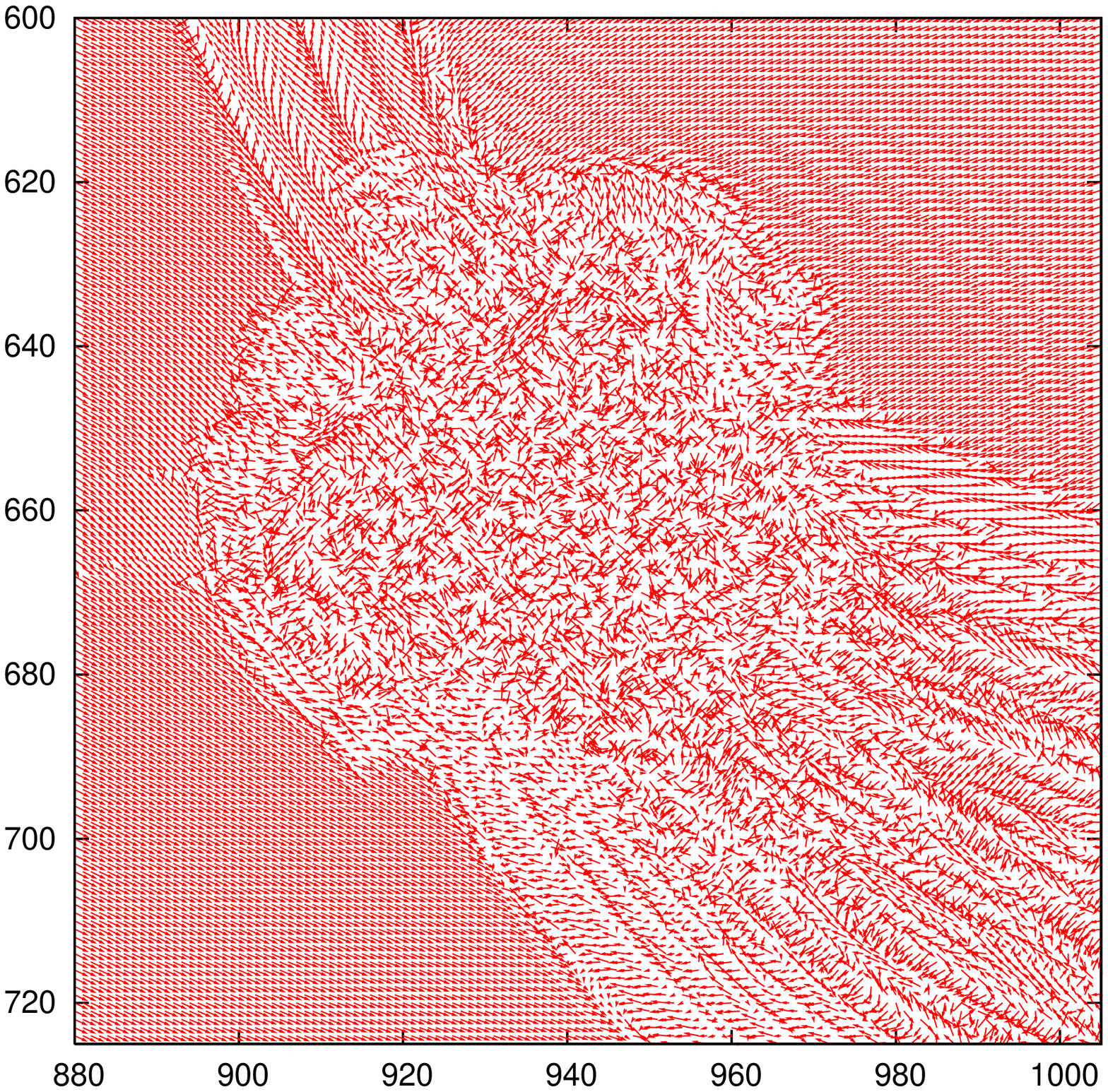}\\
  \caption{The same two-dimensional slice of the velocity field for halo B in the comoving frame.}\label{fig:Velocity_Field_Halo_B}
 \end{center}
\end{figure}

Note that the final collapsed core contains angular momentum through angular dependence in the wave function.  The angular dependence manifests itself with large-amplitude, small-scale fluctuations in the wave function.  To examine this aspect of the halo, we let the wave function be represented by $\psi= f e^{iS}$.  The specific kinetic energy is obtained through the real part of the expression:
\begin{equation}
-{{\psi^*\nabla^2 \psi}\over{2{\eta}^2 {|\psi}|^2}} = {{1}\over {\eta}^2}[(\frac{(\nabla S)^2}{2} - \frac{\nabla^2 f}{2f} ) - i (\frac{\nabla f}{f} \cdot \nabla S   +  \frac{\nabla^2 S}{2})].
\end{equation}
On the other hand, the specific flow energy can be evaluated through
the real part of the following:
\begin{equation}
-\frac{\nabla^2 ({\psi^2}/|{\psi}|^2)}{4{\eta}^2 ({\psi^2}/|{\psi}|^2))} = {{1}\over {\eta}^2} [ \frac{(\nabla S)^2}{2} - i (\frac{\nabla^2 S}{2})].
\end{equation}
Combining the two, the specific internal energy is obtained.  Plotted in Figs.(11) and (13) are the 2D slices of density for halos A and B, showing large-amplitude, small-scale fluctuations in density.  We also plot the same slice for the 2D flow velocity $\nabla_\perp S/\eta$ $(=-i\nabla_\perp(\psi^2/|\psi|^{2})/(2\eta\psi^2/|\psi|^2)$ in Figs.(12) and (14).  The velocity patterns clearly reveal well-defined boundaries of turbulence regions in halos A and B against the infall.  The flow becomes randomized inside this sharp boundary.  Despite the boundary outlines an accretion shock-like structure, we find in the density slice that there is no obvious jump at the boundary and it is not a shock.
Thus, there is no analogy of such a structure in the fluid system.  This peculiar feature warrants our further investigation in the future.

\begin{figure}
 \begin{center}
 \leavevmode
  \includegraphics[width=8cm,angle=-90]{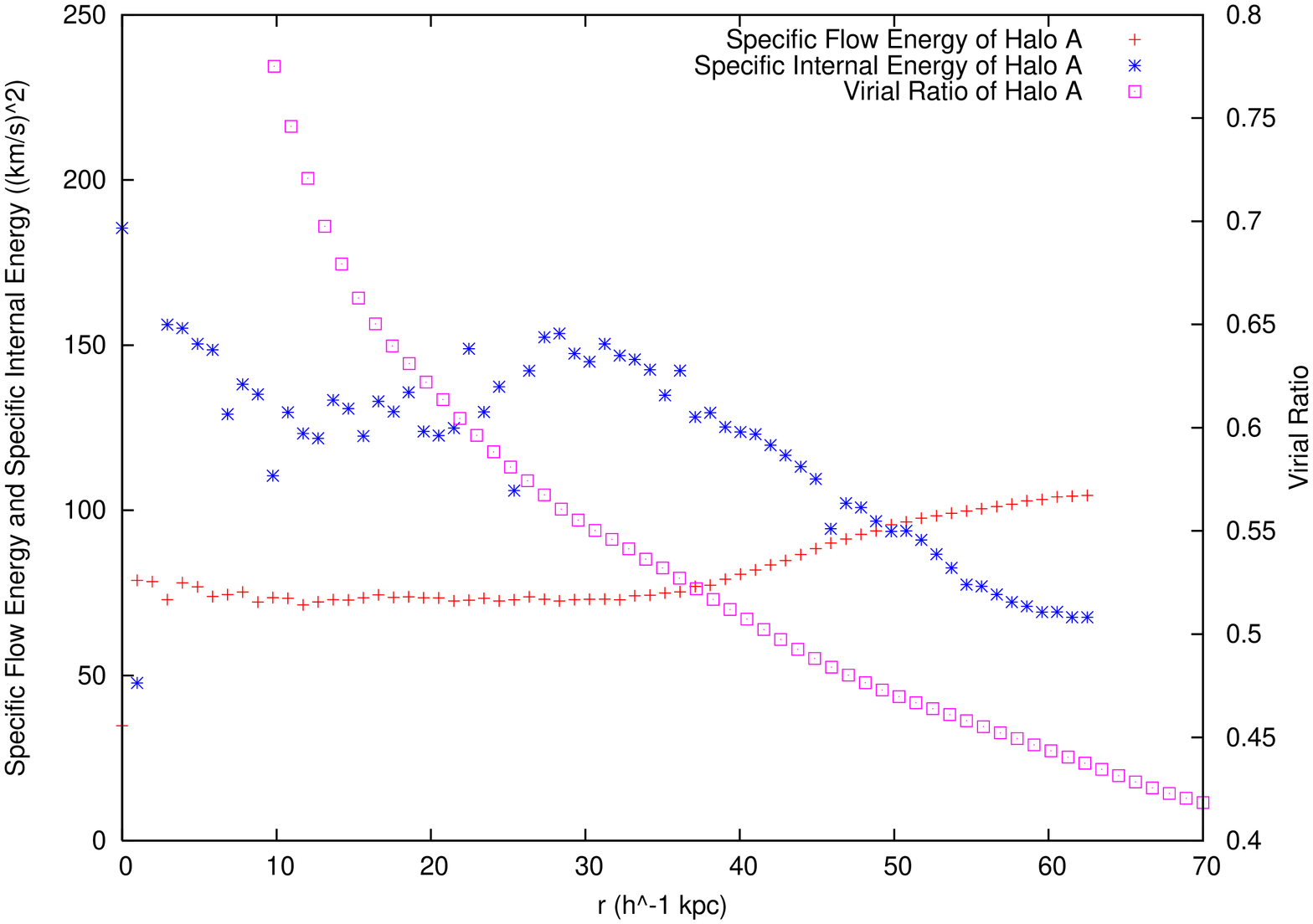}\\
  \vspace{1cm}
  \includegraphics[width=8.1cm,angle=-90]{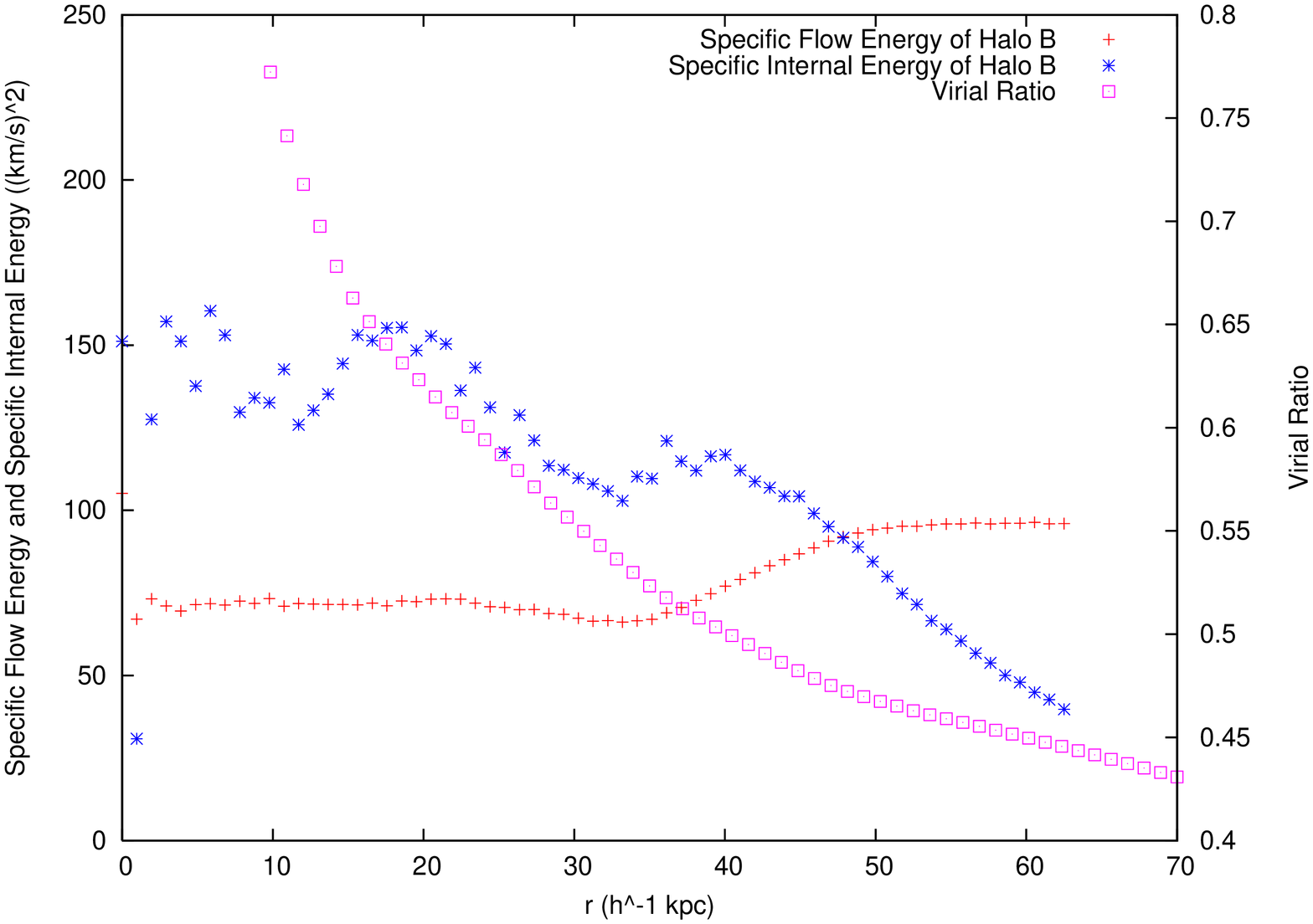}\\
  \caption{Virial ratios of  the kinetic energy integrated up to a radius $r$ (square) to the potential energy integrated up to $r$, the specific flow energy (plus) and specific internal energy (star) for halos A (upper panel) and B (lower panel). This virial ratios are about 0.5 at the average radii of the infall boundaries.
The specific internal energy is about twice as large as the specific flow kinetic energy in the interiors of the two halos.}\label{fig:Energy_K_P}
 \end{center}
\end{figure}

\begin{figure}
 \begin{center}
 \leavevmode
  \includegraphics[width=8cm,angle=-90]{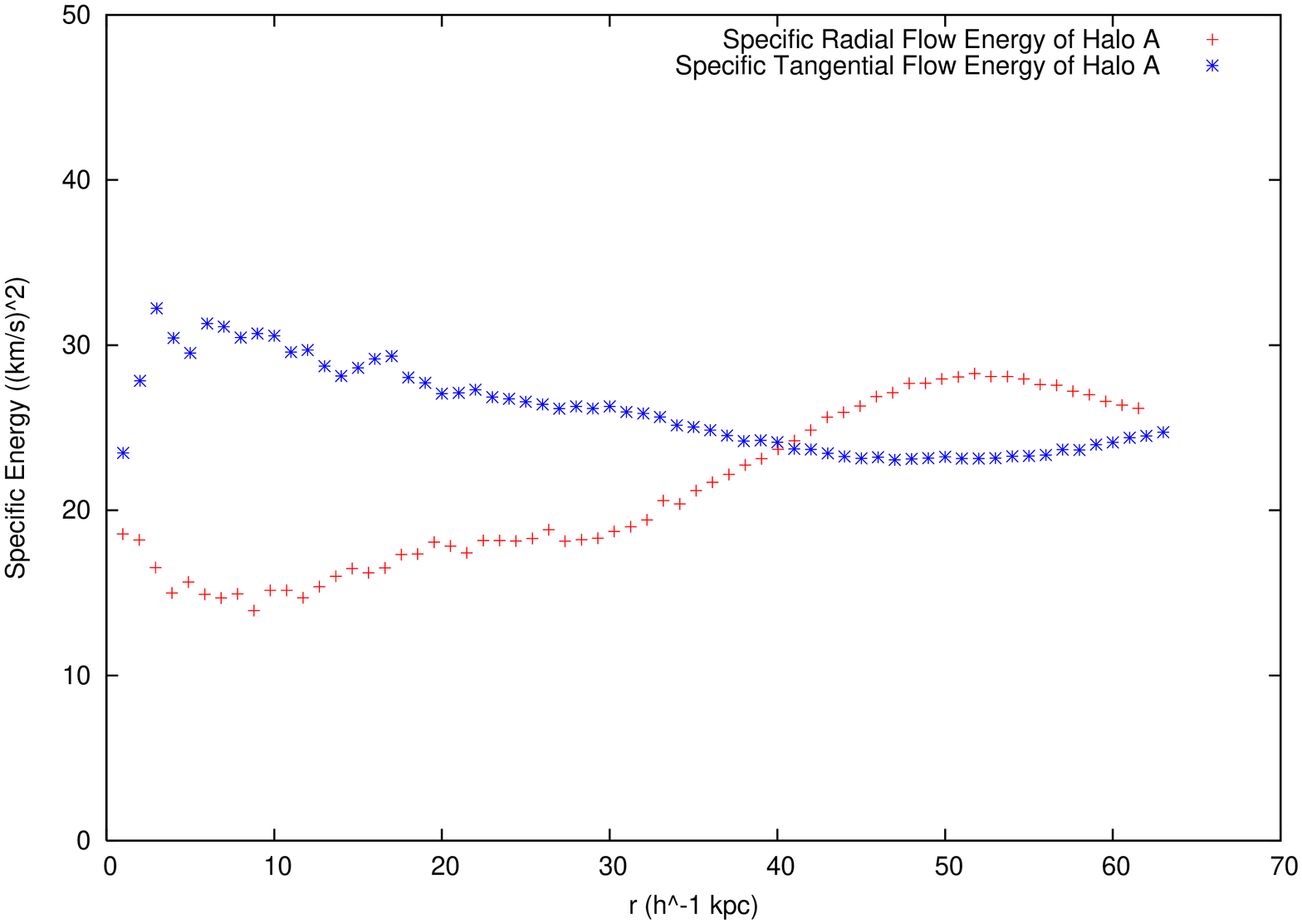}\\
  \vspace{1cm}
  \includegraphics[width=8cm,angle=-90]{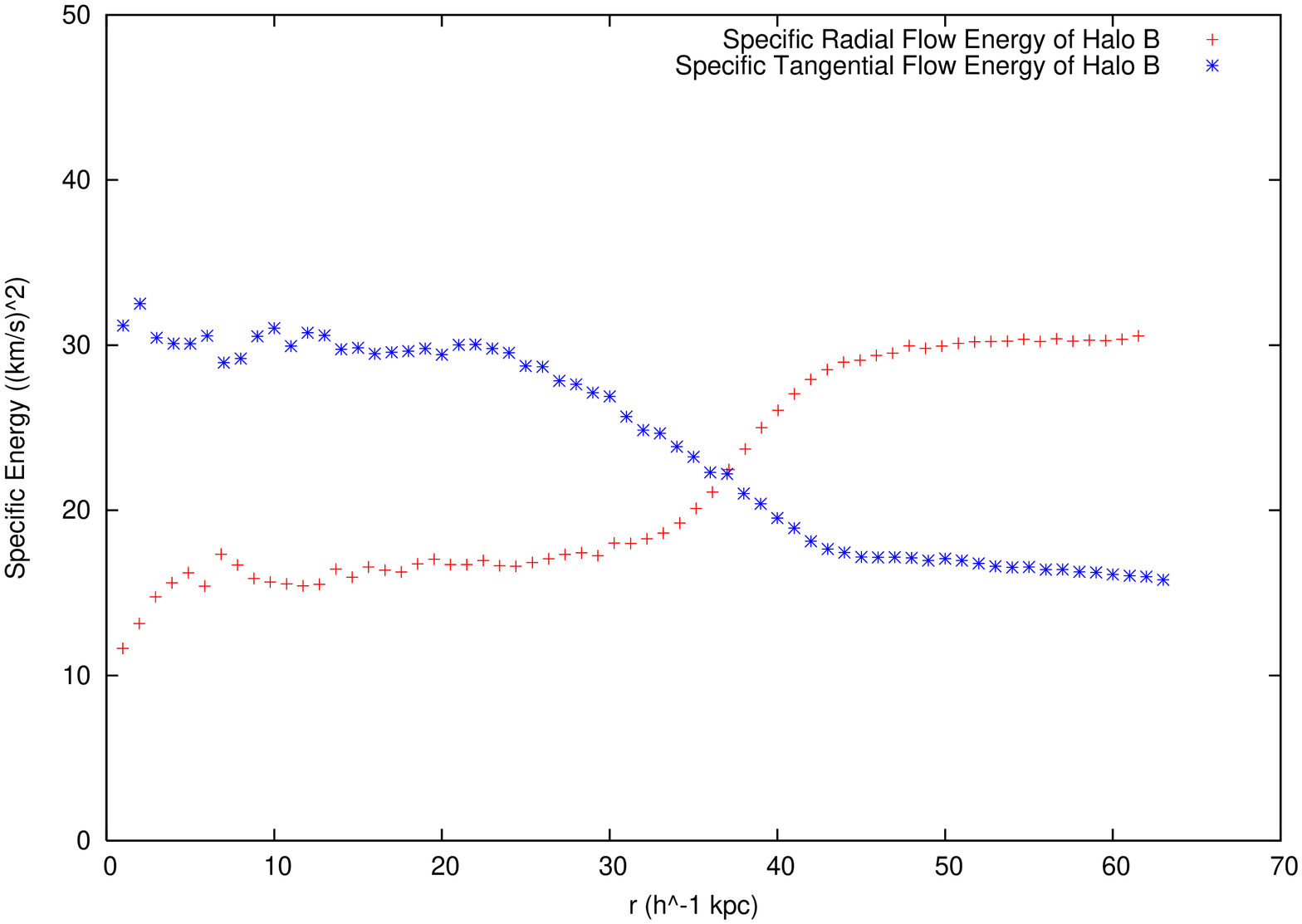}\\
  \caption{Specific radial flow energy and tangential flow energy of  halo A (upper panel) and halo B (lower panel).}\label{fig:KE_r_t}
 \end{center}
\end{figure}
We next investigate the virial conditions in the two halos.
Plotted in Figs.(15a) and (15b) are the ratios of the kinetic energy integrated up to a radius $r$ to  the potential energy
($\int_0^r 4\pi r'^2 (3\Omega_m\eta/4)[n(U-U_{min})] dr'$)
integrated up to $r$ for halos A and B.
The virial ratios are 0.5 at the average radii of the infall boundaries, within which turbulence occurs.  In addition, we also plot the specific flow energy and the specific internal energy respectively in Fig.(15). It is found that the specific internal energy is twice as large as the specific flow kinetic energy in the interiors of two halos.

Virialization can be correlated with the flow equi-partition.  Plotted in Fig.(16) are the random tangential specific flow energy and the random radial specific flow (subtracted off the mean radial infall) energy averaged over spherical shells for the two halos. The tangential flow energy is about twice the radial flow energy only well within the halos, thus providing an evidence of equi-partition at the halo cores.  In the outskirts of the halos, the random radial flow energy is larger than the equi-partition value.  This aspect is reminiscent of the velocity dispersion in a standard CDM halo.

Equi-partition is also related to the sphericity of mass distribution.  Significant large-scale angular dependence in the wave function can yield aspherical halos.  We define the quadrapole-to-monopole ratio as
\begin{equation}
Q\equiv (\frac{(\lambda_1-\lambda_2)^2+(\lambda_2-\lambda_3)^2+(\lambda_3-\lambda_1)^2}{2(\lambda_1^2+\lambda_2^2+\lambda_3^2)})^{\frac{1}{2}},
\end{equation}
where $\lambda_1$, $\lambda_2$ and $\lambda_3$ are the eigenvalues of $\int ({\bf{r}}{\bf{r}}/r^{\beta}) n d^3{\bf{r}}$, with $\beta=3.5$ to weigh in favor of the core.
The $Q$ value characterizes the low-order angular dependence of wave function, which assumes the extreme value zero or unity when the density profile has spherical symmetry or a one-dimensional shape.  It is found that $Q=0.34$ for halo A and $0.11$ for halo B respectively.  Perhaps violent relaxation after a major merger accelerates halo B to assume spherical symmetry.  We note that the quantum stress is in fact anisotropic: $T_{ij}^{Q}=(\partial_i \sqrt{n})(\partial_j \sqrt{n})/{\eta}^2 - \delta_{ij}(\nabla^2 n)/4{\eta}^2$ \citep{chiu00}.
Unlike fluid dark matter \citep{yoshida00}, the density asphericity arises from the anisotropic stress of quantum mechanics, similar to that produced by collisionless dark matter.  The similarity between the quantum dynamics and the collisionless particle dynamics in fact motivated \citet{wk93} to propose a model that approximates the latter by the former.

\section{Conclusion}

As far as we know of, this work presents the first result for the study of Bose-Einstein condensate under self-gravity via high resolution ($1024^{3}$ grids) simulation.
\citet{hu00} conjectured that if the dark matter is ELBDM, it can solve the long standing problem of far too many low-mass halos present in the standard CDM simulations, and also explains the existence of flat cores in some galaxies. In this work, we confirm that low mass halos are indeed suppressed by quantum stress even when the small-scale fluctuations are abundant in the initial power spectrum. This result is a consequence of long-time linear suppression of the
small-scale modes. We also find, from our simulations of different grid resolutions, that collapsed halos develop singular cores regardless of the halo formation processes. All these runs produce convergent density profiles.
Our $1024^{3}$ highest resolution run gives singular density profiles similar to what standard CDM simulations produce.

In retrospect this singular-core result may not be too surprising, as it arises from an almost scale-free Schr\"{o}dinger-Poisson system.
This system is not exactly scale free because there exists a Jeans length for fluctuations that are small in amplitude.  However, when the local density much exceeds the background density, the latter becomes locally ill-defined, the Jeans length no longer has any physical significance, and the Schr\"{o}dinger-Poisson system becomes locally scale free.
Being locally scale-free, the system develops singularities within a finite time.  By contrast, the conservation of phase-space density in classical particle dynamics precludes the space density of standard dark matter particles from developing any singularity \citep{chiu97,dal01}, and explains the existence of a flat core in the warm dark matter model.  Note that such a phase-space constraint does not exist for nonlinear wave dynamics; one example of this nature is a system described by the nonlinear Schr\"{o}dinger equation with attractive
self-interaction \citep{sulem99}.

Most recent observations of rotation curve in low-surface-brightness galaxies indicate inconclusive results, as far as the existence of singular halo core is concerned. Some galaxies are claimed not to possess singular halo cores,
and some are if non-circular motion is taken into consideration. Among those that do, many possess concentration parameters inconsistent with the constraint given by
$\Lambda$CDM cosmology \citep{swt03,zaku06,kuz06,kuz08}. Given the present status of observations, if galaxies indeed contain singular cores, ELBDM will likely be the only viable candidate for the dark matter that, on one hand, permits the galactic-scale, NFW-like halo cores, and on the other hand suppresses the sub-galactic low-mass halos.

\acknowledgments

\emph{Acknowledgements}- We thank Prof. Ue-Li, Pen who made the \emph{Sunnyvale cluster} of CITA available to us. We also thank Prof. Yih-Yu, Chen and Mr. Shing-Kwang, Wong for helpful discussions. This project is supported in the part by National Science Council of Taiwan under the grant: NSC 97-2628-M-002-008-MY3, and also by National Center for High-performance Computing for the availability of IBM 1350.

\appendix

\begin{center}
  {\bf APPENDIX}
\end{center}

\renewcommand\thefigure{{A}.\arabic{figure}}
\setcounter{figure}{0}

\section{Comparison with One and Two Dimensional Simulations:}

We also examine the one-dimensional and two dimensional simulations
with the same numerical scheme used for the three-dimensional
simulations.  It is hoped that we can verify the scheme by reproducing
the results of previous works.

So far as we know of, the only work on one-dimensional simulation of the same
physical system is the work of \citet{hu00}.  In this work, they simulated
the 1D sheet collapse in a three-dimensional expanding universe.  Two cases of
different Jeans lengths were examined.  For the short Jeans length case,
they found fast oscillating solutions, in time and space, which corresponds
to a solution of repeated collapses and rebounds.   For the long Jeans
length case, they found slow oscillating solution.  In either case, the
collapsed sheet cores are smooth with no cusp.

The only two-dimensional simulation in the literature is that of \citet{wk93}.
Unfortunately, they did not examine individual profiles of
collapsed filaments and no comparison can be made, since the focus of
this work placed on the equivalence of the light bosonic system and
the cold dark matter system.

We run one-dimensional simulation with 8192 grids and a small-amplitude
initial random fluctuation.   The evolution follows the linear growth
described in the text.  Nonlinearity sets in quite quickly when the
density fluctuation grows to a amplitude comparable to the background density.
The saturation amplitude of $\delta\rho/\rho$ is around 10.  The solution
indeed oscillates, corresponding to repeated collapses
and rebounds.   We show the solutions in Fig.(A1) for two nearby
time steps where $\Delta a=0.001$ near $a=1$.  It is apparent that the
solution has no singular core.  This solution is similar to what
was found in the work of \citet{hu00}.

\begin{figure}
 \begin{center}
 \leavevmode
  \includegraphics[width=14cm,angle=0]{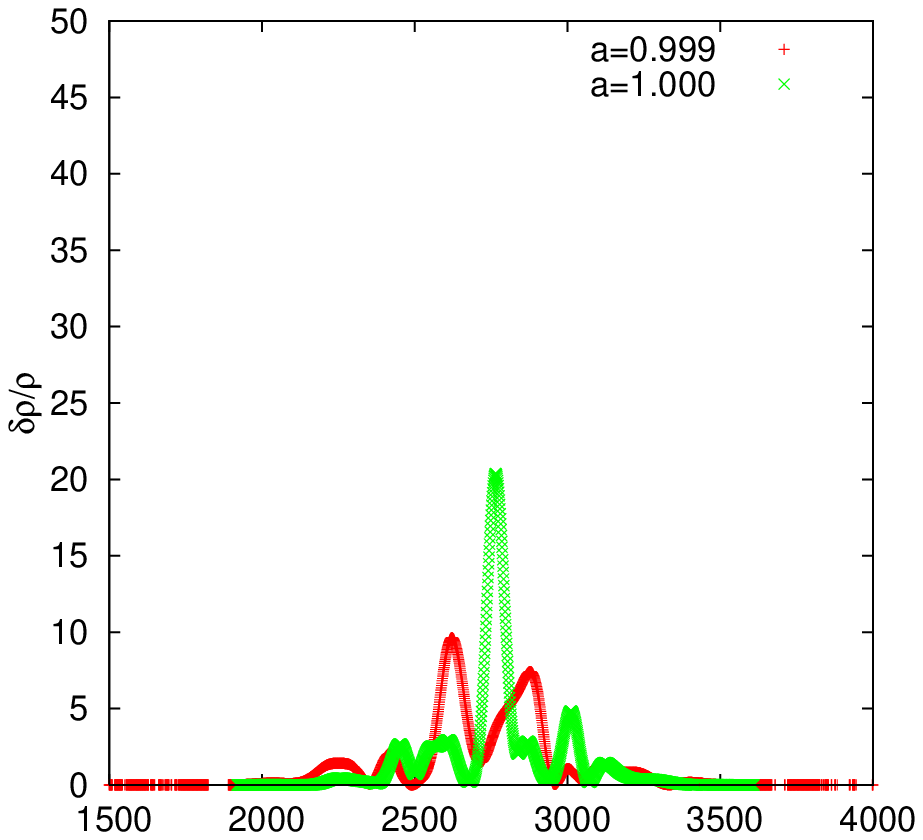}\\
  \caption{The density profiles of two nearby
time steps where $\Delta a=0.001$ near $a=1$ in a 1D simulation of 8192 grids.}\label{fig:Hu_result}
 \end{center}
\end{figure}

The two-dimensional case has also been examined with the same scheme as the 3D
code with 4096$^2$ grids.  We find that the collapsed filaments at $z=0$ do
not contain singular cores; rather, several dense clumps of
$\delta\rho/\rho\sim 1000$ are scattered over a area of finite radius in the
core region, as shown in the zoom-in image in Fig.(A2).

\begin{figure}
 \begin{center}
 \leavevmode
  \includegraphics[width=14cm,angle=0]{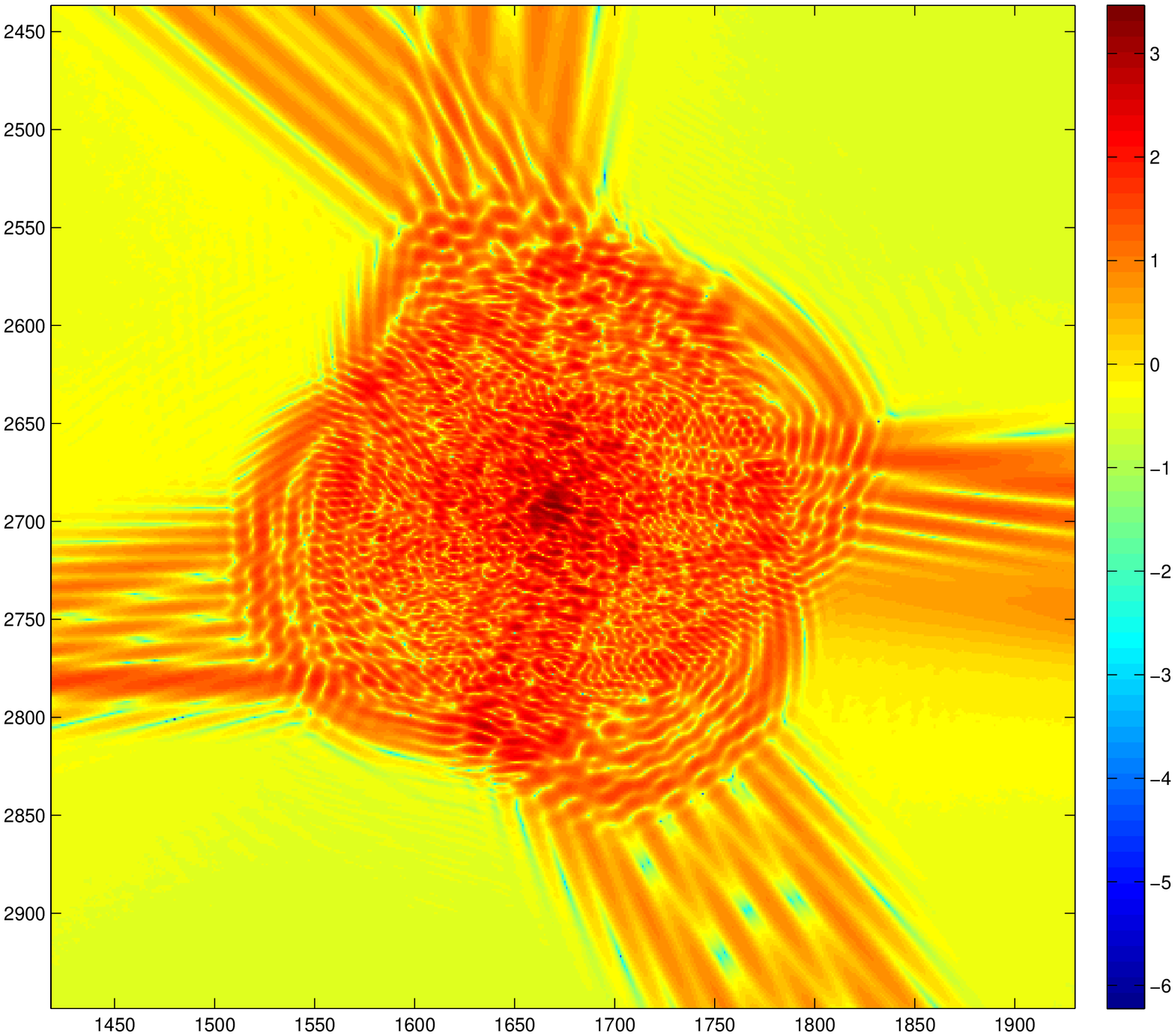}\\
  \caption{The density of the biggest halo found at the z=0 snapshot in
the 2D simulation of $4096^2$ grids. In this plot, the collapsed filament
does not contain any singular core.}\label{fig:2D_result}
 \end{center}
\end{figure}

It is understandable why low dimensional objects do not develop singularities.
This is due primarily to weakening of the focusing power in a
three-dimensionally expanding universe.  It appears that 2D is the critical
dimension, where the singularities are just not to appear.  In 3D, the
self-focusing power can be so strong as to bind the dense clumps tightly to
form singularity.

\end{document}